\documentclass[twocolumn,times]{aastex63}

\usepackage{graphicx}
\usepackage{amsfonts,amsmath,amssymb}
\usepackage{xspace}
\usepackage{bm,booktabs}
\usepackage{longtable,topcapt,rotating }

\newcommand{\snowglobes}{SNOwGLoBES}

\newcommand{\alphL}{\ensuremath{\alpha_\Lambda}\xspace}

\received{\today}
\submitjournal{ApJ}

\shorttitle{Constraining properties of CCSNe with multi-messenger signals}
\shortauthors{Warren et al.}

\begin{document}

\title{Constraining properties of the next nearby core-collapse supernova with multi-messenger signals}

\author[0000-0001-9440-6017]{MacKenzie L.~Warren}
\altaffiliation{NSF Astronomy \& Astrophysics Postdoctoral Fellow}
\affiliation{Department of Physics, North Carolina State University, Raleigh, NC 27695, USA}
\affiliation{Department of Physics and Astronomy, Michigan State University, East Lansing, MI 48824, USA}
\affiliation{Joint Institute for Nuclear Astrophysics-Center for the Evolution of the Elements, Michigan State University, East Lansing, MI 48824, USA}

\author[0000-0002-5080-5996]{Sean M.~Couch}
\affiliation{Department of Physics and Astronomy, Michigan State University, East Lansing, MI 48824, USA}
\affiliation{Joint Institute for Nuclear Astrophysics-Center for the Evolution of the Elements, Michigan State University, East Lansing, MI 48824, USA}
\affiliation{Department of Computational Mathematics, Science, and Engineering, Michigan State University, East Lansing, MI 48824, USA}
\affiliation{National Superconducting Cyclotron Laboratory, Michigan State University, East Lansing, MI 48824, USA}

\author[0000-0002-8228-796X]{Evan P.~O'Connor}
\affiliation{Department of Astronomy and Oskar Klein Centre, Stockholm
  University, AlbaNova, SE-106 91 Stockholm, Sweden}

  \author[0000-0001-6806-0673]{Viktoriya Morozova}
 \affiliation{Department of Physics, The Pennsylvania State University, University Park, PA 16802, USA}

  \correspondingauthor{MacKenzie Warren}
  \email{mlwarre4@ncsu.edu}

  \begin{abstract}
With the advent of modern neutrino and gravitational wave detectors, the promise of multi-messenger  detections  of  the  next  galactic core-collapse  supernova  has  become  very real.  Such detections will give insight into the core-collapse supernova mechanism, the structure of the progenitor star,  and may  resolve longstanding questions in fundamental physics.  In order to properly interpret these detections, a thorough understanding of the landscape of possible core-collapse supernova events, and their multi-messenger signals, is needed.  We present detailed predictions of neutrino and gravitational wave signals from 1D simulations of stellar core collapse, spanning the landscape of core-collapse progenitors from $9-120\,\mathrm{M}_{\odot}$.  In order to achieve explosions in 1D, we use the STIR model, which includes the effects of turbulence and convection in 1D supernova simulations to mimic the 3D explosion mechanism.  We study the gravitational wave emission from the 1D simulations using an astroseismology analysis of the proto-neutron star.   We find that the neutrino and gravitational wave signals are strongly correlated with the structure of the progenitor star and remnant compact object.  Using these correlations, future detections of the first few seconds of neutrino and gravitational wave emission from a galactic core-collapse supernova may be able to provide constraints on stellar evolution independent of pre-explosion imaging and the mass of the compact object remnant prior to fallback accretion.  

  \end{abstract}


\section{Introduction\label{sec:intro}}

With the proven success of modern neutrino and gravitational wave (GW) detectors, we have entered the era of truly multi-messenger astronomy.  Perhaps one of the most exciting prospects for multi-messenger astronomy are core-collapse supernovae (CCSNe).  CCSNe result from the deaths of massive stars.  Stars with ZAMS masses $M\gtrsim 8 \,\mathrm{M}_{\odot}$ will undergo gravitational collapse following their nuclear burning lifetimes.  Once massive stars generate iron cores more massive than the effective Chandrasekhar mass limit \citep{baron:1990}, pressure support is insufficient to balance gravity and the core begins to collapse. As the core collapses, electron captures generate a tremendous number of electron neutrinos. The inner core of the star rebounds, sending pressure waves through the infalling material and forming an outward moving shock.  After core bounce, thermal processes create neutrinos and antineutrinos of all flavors. Ultimately $\sim99\%$ of the gravitational binding energy of the proto-neutron star (PNS) ($\sim \mathrm{a \,few\,} \times10^{53}$ ergs) is converted into neutrinos and results in a dense neutrino field.  As the shock moves through the outer mantle of the iron core, it quickly loses energy to the dissociation of heavy nuclei into free protons and neutrons and soon stalls, becoming an accretion shock.  

Although the details of reviving the stalled shock in order to generate a successful explosion are still not fully resolved \citep{bethe:1990, janka:2007, janka:2012a, janka:2016, janka:2012, burrows:2013,melson:2015,lentz:2015,takiwaki:2016, muller:2016b, couch:2017,ott:2018,oconnor:2018,cabezon:2018,burrows:2019}, it has become clear that the stalled shock will not be successful in all cases.  Some fraction of massive stars will result in ``failed'' supernovae \citep{oconnor:2011,lovegrove:2013,ertl:2016,sukhbold:2016,adams:2017,sukhbold:2018,couch:2020} -- massive stars that go through core-collapse and bounce, only to be unable to generate a successful explosion and resulting in BH formation. Although such an event may or may not result in an electromagnetic transient \citep{lovegrove:2013,adams:2017,quataert:2019}, ``failed'' supernovae may have distinct neutrino and GW emission that will allow these events to be detected and distinguished from ``successful'' supernovae (e.g.~\cite{sumiyoshi:2006}).

Stellar core collapse events are prime candidates for multi-messenger astronomy: each CCSN event will produce prodigious numbers of neutrinos ( $\gtrsim10^{57}$).  To date, supernova neutrinos have only been detected for one event: SN1987A, from which $\sim25$ neutrinos were detected over $\sim10 \,\mathrm{s}$ \citep{arnett:1989}.   Such an event will be detectable with modern neutrino detectors if it occurs in our galaxy \citep{scholberg:2012,suwa:2019}, and potentially beyond with planned future detectors such as Hyper Kamiokande \citep{scholberg:2012,himmel:2016,HyperK}. aLIGO, Virgo, and KAGRA will only be able to detect nearby CCSNe, at a range of $\leq 100\,\mathrm{kpc}$ \citep{abbott2016}.  GWs from a CCSN have not yet been observed. 

The combination of observing neutrinos and GWs will provide multiple views into a single CCSN event.  Although ideally GWs, neutrinos, and electromagnetic emission will be detected from a single CCSN event, the question remains how much information can be extracted from detection via just one messenger.  For example, a CCSN event that occurs in the plane of the galaxy could be entirely obscured from electromagnetic (EM) detection, but will still be detectable in neutrinos and, potentially, GWs.  Will we be able to distinguish a ``successful'' versus ``failed'' CCSN event without direct detection of EM emission?  What can we tell about the progenitor star, explosion mechanism, and remnant compact object from detection of neutrinos, GWs, or both? 

Additionally, as the sensitivities and observation limits of neutrino and GW detectors are improved in coming decades, there is hope of detecting the diffuse supernova neutrino background \citep{moller:2018,horiuchi:2018,hidaka:2018,mathews:2019} or GW background \citep{buonanno:2005,kotake:2007,crocker:2017}.  However, providing predictions of these backgrounds requires a detailed understanding of the outcomes of CCSN events for all stars with masses and metallicities relevant for the span of cosmic history, as well as reliable predictions of the neutrino and GW emission from such a diverse spectrum of events.

3D CCSN simulations will be limited in their use for addressing these questions for the foreseeable future.  Such simulations are too computationally expensive to be utilized in sufficiently large sensitivity studies \citep{nagakura:2019}.
 Spherically symmetric CCSN simulations remain vital to our understanding of CCSN physics and the contributions of CCSNe to galactic chemical evolution.  The relatively small computational cost of 1D simulations and the ability to run simulations to late times lends them to these sorts of studies (e.g.,~progenitor mass \citep{ugliano:2012,perego:2015,pejcha:2015,ertl:2016,sukhbold:2016,ebinger:2019,sukhbold:2018,couch:2020} or to study nucleosynthesis at late times \citep{frohlich:2006a,frohlich:2006b,frohlich:2006}).

In this work, we present a study of multi-messenger signals -- GW and neutrino -- from the landscape of massive stellar core collapse  using 1D simulations.  We utilize the 200 single star, solar metallicity, non-rotating progenitors of \cite{sukhbold:2016} with zero-age main sequence (ZAMS) masses in the range $9-120\, \mathrm{M}_{\odot}$. Our results are the first detailed predictions for the observable GW and neutrino signals covering the vast majority of the landscape of iron core, single star CCSN progenitors.\footnote{Publicly available at \url{https://doi.org/10.5281/zenodo.3667909}}

Additionally, with a large set of simulations, we are able to \textit{statistically} study the sensitivity of observable neutrino and GW signals to fundamental CCSN physics, as well as correlations between observable signals and progenitor, in a robust manner.   Previous works have explored temporal and spatial correlations between the neutrino and GW signals for single CCSN simulations, largely with the aim of measuring the presence of SASI and similar phenomena \citep{fuller:2015,camelio:2017,kuroda:2017,takiwaki:2018,vartanyan:2019}.  For rapidly rotating CCSNe, \cite{schneider:2019} found that the characteristic frequency of the gravitational wave signal can be imprinted on the neutrino signal and may be detectable for very nearby CCSNe with current facilities such as IceCube.  

\cite{torres-forne}, \cite{sotani:2016}, \cite{sotani:2019}, and \cite{mueller:2013} found relationships between the GW emission and properties of the progenitor, explosion, and remnant compact object.  However, these works were limited to a few progenitor stars and only considered the GW emission.   \cite{richers:2017} explored correlations between equation of state parameters with the gravitational wave signal at bounce for a variety of nuclear equations of states for rotating CCSNe and \cite{sotani:2017} found for a single progenitor that detection of two or more gravitational wave modes could be used to constrain the nuclear equation of state. 

Similarly, there are many works that have studied relationships between the neutrino signal and progenitor and explosion properties.  \cite{suwa:2019} proposed that the late time neutrino emission can be used to constrain the PNS mass for successful CCSN events.  \cite{nakamura:2015} used several hundred 2D CCSN simulations to explore correlations between the neutrino luminosity, explosion energy, and progenitor and \cite{horiuchi:2017} used 2D simulations to determine the relationship between the neutrino signal and core-compactness. For a review of constraints from the detection of CCSN neutrinos, we point the reader to \cite{horiuchi:2018}.

Here, we use correlated neutrino and GW signals to determine the progenitor mass, compact object mass, explosion energy, etc, for \textit{any} possible CCSN event.  Additionally, by considering \textit{multi-messenger} signals in our analysis, we find signals from different messenger sources that may be combined to determine properties of the progenitor star and explosion with more certainty than using a single messenger alone.  In Section~\ref{sec:methods} we discuss the various codes that we use to simulate the CCSNe and resultant neutrino and GW  emission.  Our results are laid out in Section~\ref{sec:results}: in Section~\ref{sec:expld}, we discuss the observable differences between failed and successful core collapse events in neutrinos and GWs.  The correlations between progenitor and explosion properties and neutrino emission is discussed in Section~\ref{sec:neutrino}, and the GW emission in Section~\ref{sec:gw}.  In Section~\ref{sec:corr}, we consider how signals from different messenger sources may be combined to obtain progenitor properties with increased certainty. We provide our discussion and conclusions in Section~\ref{sec:conclusions}.

\section{Methods\label{sec:methods}}
To conduct this study, we start with a massive progenitor star, evolved to the point of core collapse.  We then simulate the core collapse event itself, and resultant explosion (or failure).  We use the FLASH multiphysics simulation framework \citep{fryxell:2000}, the details of which are laid out in Section~\ref{sec:flash}, to conduct a series of 1D CCSN simulations using the Supernova Turbulence In Reduced-dimensionality (STIR) model \citep{couch:2020}.  We simulate the expected neutrino signal as would be observed in a Super Kamiokande-like detector using the \snowglobes  $\,$code \citep{scholberg:2012}, as detailed in Section~\ref{sec:methods-snow}, and calculate the expected GW signal using an astroseismology analysis as discussed in Section~\ref{sec:methods-gw}.  Finally, in Section~\ref{sec:methods-corr} we present our framework for analyzing the relationships between underlying progenitor and explosion properties with the observable signals.

We use the 200 solar metallicity progenitor models provided by \cite{sukhbold:2016}, with ZAMS masses ranging from $9-120 \,\mathrm{M}_{\odot}$.  These stellar models were created with the KEPLER stellar evolution code, assuming zero rotation and magnetic field and single star evolution.  We direct the reader to the works of \cite{sukhbold:2016}, and references therein, for the details of these models.

\subsection{FLASH\label{sec:flash}}

We simulated the CCSN events using the {FLASH} code with the STIR (Supernova Turbulence In Reduced-dimensionality) turbulence-driven explosion model \citep{couch:2020}, a new method for artificially driving CCSN explosions in 1D simulations. Turbulence is crucial for understanding the CCSN explosion mechanism, since turbulence may add $>$20\% to the total pressure behind the shock \citep{couch:2015a} while turbulent dissipation contributes significantly to post-shock heating \citep{mabanta:2018}, thus aiding in the explosion. Including turbulence results in successful explosions in 1D simulations, replicating the current understanding of the physical explosion mechanism, and reproduces the thermodynamics and composition seen in 3D simulations \citep{couch:2020}.

STIR uses the Reynolds-averaged Euler equations with mixing length theory as a closure to model the effects of turbulence.  
This model has one primary free parameter, the mixing length parameter, $\alpha_\Lambda$, which is the usual scaling parameter in mixing length theory \citep[cf.][]{cox:1968} and scales the strength of the convection.  The parameter has been fit to 3D simulations run with FLASH. For details of the STIR model and the fitting to 3D simulations, we point the reader to \cite{couch:2020}.

    Recently, \citet{mueller:2019} has criticized methods for including turbulence in 1D simulations such as STIR and that of \citet{mabanta:2019} for not explicitly enforcing total energy conservation. 
    M\"uller's principle critique is that the neglect of the turbulent mass flux term in the Reynolds-averaged continuity equation in both STIR and the model of \citet{mabanta:2019} breaks total energy non-conservation.  
    As discussed by \citet{murphy:2011} and demonstrated in \citet{murphy:2013}, this term is second-order in the perturbation and thus expected to be a small correction to the overall mass flux. 
    We have verified this quantitatively in our STIR models finding that the neglected turbulent mass flux is no larger than about 1\% of the background mass flux at any time during our simulations.
    So, while it is true that precise energy conservation would require inclusion of this turbulent mass flux in the continuity equation, the inclusion of diffusive mixing of energy and composition serves to approximately conserve total energy instead.
    As explicitly discussed by \citet{couch:2020}, and acknowledged by \citet{mueller:2019}, the turbulent energy in STIR is extracted from the free energy in unstable entropy and composition gradients that drive the buoyant convection and turbulence.
    Therefore, while strict energy conservation is not ensured in STIR, it is approximately conserved depending on the values of the diffusive mixing parameters. 
    Furthermore, all models for driving CCSN explosions in 1D \citep[e.g.,][]{ugliano:2012,perego:2015,ertl:2016,sukhbold:2016,mueller:2016} must make some abuse to physics which often can be characterized as implying a non-conservation of energy. 
    Take, for instance, the model used by \citet{ugliano:2012, ertl:2016, sukhbold:2016}. 
    Here, 1D explosions are achieved by enforcing a rapid contraction of the PNS in order to enhance the emergent neutrino luminosities, yielding explosions. 
    Thus, while some integral resembling the total energy could be constructed and shown to be conserved in this model, the ad hoc prescription for the contraction of the PNS is manifestly inconsistent with the chosen treatment of gravity and equation of state in this model, implying that total energy is not truly conserved. 
    
    The purpose of such 1D models is after all to \textit{approximate} the dynamics of full 3D CCSN simulations, but far more quickly and cheaply.
    To this end, the true test of the validity of a 1D model should be its ability to reproduce the results of comparable 3D simulations, while making reasonable approximations to the physics of the problem.
    In \citet{couch:2020}, the authors compare the results of STIR to those of \citet{burrows:2019}, the largest collection of high-fidelity 3D CCSN simulations in the literature at present. 
    In {\it every single} case of the same progenitor model, STIR correctly predicts the key result of explosion or failure as compared with the 3D simulations of \citet{burrows:2019}. 
    As shown in \citet{couch:2020}, other 1D explosion models such as that of, e.g., \citet{ebinger:2019} and \citet{ertl:2016} do not fair so well by this metric.
    Furthermore, STIR also reproduces the angle-averaged convective and thermodynamic structure of full 3D simulations closely \citep{couch:2020}.
    The argument that a parameterized 1D model for CCSN explosions must perfectly conserve energy to hold any validity is a fallacious false dilemma.

As was done in \cite{couch:2020}, we assume fiducial values for the diffusion coefficients ($\alpha_K = \alpha_{e} = \alpha_{Y_{e}} = \alpha_{\nu} = 1/6$) and let the mixing length coefficient $\alpha_{\Lambda}$ vary between 1.23 and 1.27.  $\alpha_{\Lambda} = 0.0$ would correspond with a simulation without convection and turbulence, e.g. the ``typical'' 1D simulation.  \cite{couch:2020} find that the ``best fit'' value of $\alpha_{\Lambda}$, when comparing to the convection seen in 3D simulations, is about 1.25.  Here, we allow this value to vary slightly around the best fit value also, including simulations with $\alpha_{\Lambda} = 1.23$ and $\alpha_{\Lambda} = 1.27$, to account for uncertainties in the theoretical models, uncertainties in the fitting to the 3D simulation, uncertainties in our understanding of the explosion mechanism, the 1D nature of these simulations, and the multitude of uncertainties in our understanding of stellar evolution such as binarity, rotation, etc. (see, e.g., \citealt{clausen:2015}). 

\begin{figure}[b]
   \centering
   \includegraphics[width=0.5\textwidth]{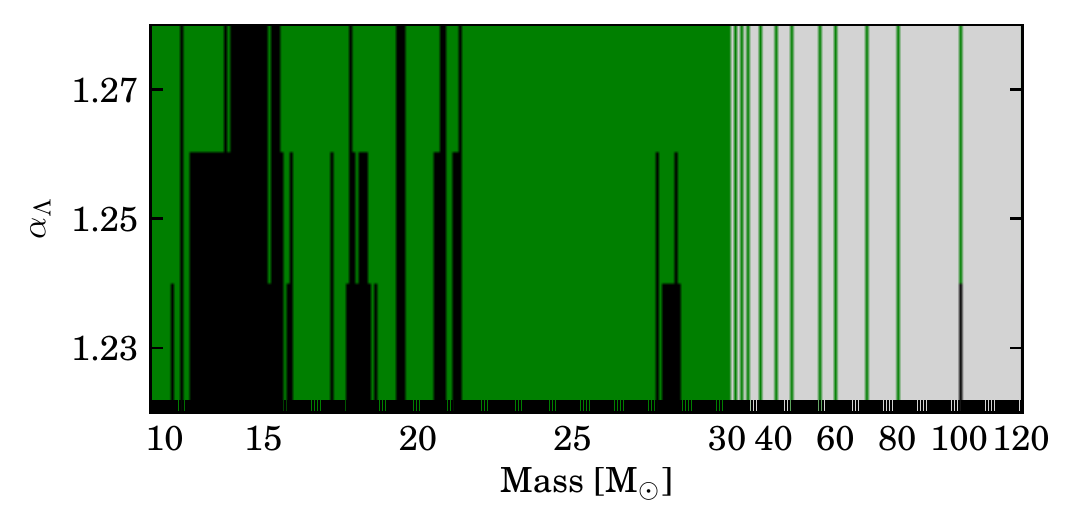} 
   \caption{Explodability for supernova progenitors from $9-120 \,\mathrm{M}_{\odot}$ for turbulence strength parameters $\alpha_{\Lambda} = 1.23 - 1.27$.  Green indicates a simulation that did explode successfully, while black indicated a progenitor that does not explode and results in a ``failed'' supernova event.  Gray regions indicated mass regions that were not simulated.  The non-monotonicity of explodability with ZAMS mass has been seen by other groups \citep{oconnor:2011,perego:2015,sukhbold:2016} and is indicative of the sensitivity of the explosion mechanism to the core structure at the time of collapse.}
   \label{fig:stir}
\end{figure}

Figure~\ref{fig:stir} shows the explodability of the progenitor stars considered in this study, using the {STIR} model.  As has been seen in many previous studies \citep{oconnor:2011,perego:2015,sukhbold:2016,couch:2020}, the explodability is not simply related to ZAMS mass.  Most notably, STIR predicts a large range of failed explosions and black hole formation around $M_\mathrm{ZAMS} = 13-15\,\mathrm{M}_\odot$.  A gap in explodability is also seen in this mass range by \cite{sukhbold:2018}, although it is narrower, and failed explosions are also predicted by 2- and 3-D simulations in this mass range as well \citep{vartanyan:2018,burrows:2019}.
This gap would appear to be in tension with the observed masses of red supergiant CCSN progenitors \citep[e.g.,][]{smartt:2009}.
However, extreme caution should be taken when consider the implications of this since there are substantial  sensitivities in the progenitor models \citep{sukhbold:2018,sukhbold:2020} and uncertainties in the ZAMS mass estimation of observationally identified CCSN progenitors \citep{farrell:2020}.
For further discussion of the landscape of explodability predicted by the STIR model, and how it compares to other codes, we direct the reader to \cite{couch:2020}. 

The simulations with STIR were run until one of three conditions was met: the shock reached the outer grid boundary ($15,000 \,\mathrm{km}$), the PNS collapsed to a BH, or a maximum simulation time of $5\,\mathrm{s}$.   In principle, there is no reason why explosions cannot occur after $5\,\mathrm{s}$ post-bounce, but we will consider simulations that reach $5\,\mathrm{s}$ without exploding or collapse to a BH as ``failed'' CCSNe.  However, for much of this work, we will not include these simulations in our study, unless otherwise noted, because we do not truly know the outcome of these simulations and do not have a reliable measure of the time to collapse to BH. 

STIR evolves the turbulent kinetic energy and incorporates corrections due to the turbulent flow directly into the hyperbolic fluxes in the hydrodynamic solver of the FLASH adaptive mesh refinement (AMR) simulation framework \citep{fryxell:2000,dubey:2009}.  The hydrodynamic solver is a newly implemented \citep{couch:2020} fifth-order finite-volume weighted essentially non-oscillatory (WENO) spatial discretization \citep{shu:1988,tchekhovskoy:2007} and a method-of-lines Runge-Kutta time integration \citep{shu:1988}.  We use an HLLC Riemann solver everywhere except at shocks, where a more diffusive HLLE solver \citep{toro:2009} is utilized.  FLASH treats self-gravity as an approximate general relativistic effective potential \citep{marek:2009,oconnor:2018}.
The simulations in this work were run with 10 levels of refinement in a radial domain that extends to $15,000\, \mathrm{km}$, thus the finest grid spacing is $0.229\, \mathrm{km}$.  The maximum allowed level of refinement is limited logarithmically with radius such that $\Delta r /r \lesssim 0.7\%$.

We use an explicit two-moment ``M1'' neutrino transport scheme with an analytic closure \citep{shibata:2011,cardall:2013,oconnor:2015}.  The details of the M1 implementation in FLASH can be found in \cite{oconnor:2018}.  FLASH assumes three neutrino flavors for the transport: $\nu_{e}$, $\bar{\nu}_{e}$, and $\nu_{x}$, where $\nu_{x}$ is a combined $\mu-\tau$ neutrino and antineutrino flavor.  The simulations in this work include velocity dependent terms and inelastic neutrino-electron scattering \citep{oconnor:2015} and 12 logarithmically spaced neutrino energy groups, ranging from $1$ to $300 \,\mathrm{MeV}$.  We use the NuLib neutrino opacity set \citep{oconnor:2015} that largely matches that of \citep{bruenn:1985}, with the addition of corrections due to weak magnetism \citep{horowitz:2002}.  We exclusively use the ``optimal'' equation of state of \cite{steiner:2013} in this work (``SFHo'') and assume nuclear statistical equilibrium everywhere in our simulations.  The choice of equation of state will most definitely impact both the GW \citep{richers:2017,pan:2018} and neutrino emission \citep{yasin:2020,aschneider:2019}, but we leave the exploration of the impact of the equation of state to future work.

\subsection{\snowglobes\label{sec:methods-snow}}

We use the {\snowglobes} code \citep{scholberg:2012} to simulate the observed neutrino signal.  {\snowglobes}\footnote{Available at \url{https://github.com/SNOwGLoBES/snowglobes}} is an open source code for calculating event rates in common neutrino detector materials.  Although FLASH produces detailed neutrino fluxes and spectra, we choose to post-process these fluxes through \snowglobes~to ensure that we are focusing on detectable signals, rather than running the risk of finding correlations and coming to conclusions that would not be feasible to detect in current or near-future neutrino detectors.  The next galactic CCSN will be observed in more than one detector and combining the detections from a multitude of detectors will give greater insight as each detector is uniquely sensitive to different energies and neutrino flavors.  For the purpose of this analysis, we restrict ourselves to Super Kamiokande. The problem of determining the ideal combination of detectors to interpret the neutrino signal is left to future work.

\begin{table}[t]
   \centering
   \begin{tabular}{@{} lcr @{}} 
      \toprule
      Channel & Reaction & Neutrino flavor \\
      \midrule
      IBD & $\bar{\nu}_{e} + p \rightarrow n + e^{-}$ & $\bar{\nu}_{e}$  \\
      ES & $\nu + e^{-} \rightarrow \nu + e^{-}$ &  $\nu_{e}$, $\bar{\nu}_{e}$, $\nu_{x}$ \\
      $\nu_{e}-^{16}\mathrm{O}$ & $\nu_{e} + ^{16}\mathrm{O}  \rightarrow e^{-} + ^{16} \mathrm{F}$ & $\nu_{e}$ \\
      $\bar{\nu}_{e}-^{16}\mathrm{O}$ & $\bar{\nu}_{e} + ^{16}\mathrm{O} \rightarrow e^{+} + ^{16}\mathrm{N}$ & $\bar{\nu}_{e}$ \\
      NC & $\nu + ^{16} \mathrm{O} \rightarrow \nu + ^{16}\mathrm{O}$ & $\nu_{e}$, $\bar{\nu}_{e}$, $\nu_{x}$ \\
      \bottomrule
   \end{tabular}
   \caption{Detection channels for a water Cherenkov detector and their flavor sensitivities.}
   \label{tab:channels}
\end{table}

\begin{figure}
\centering
\includegraphics[width = 0.5\textwidth]{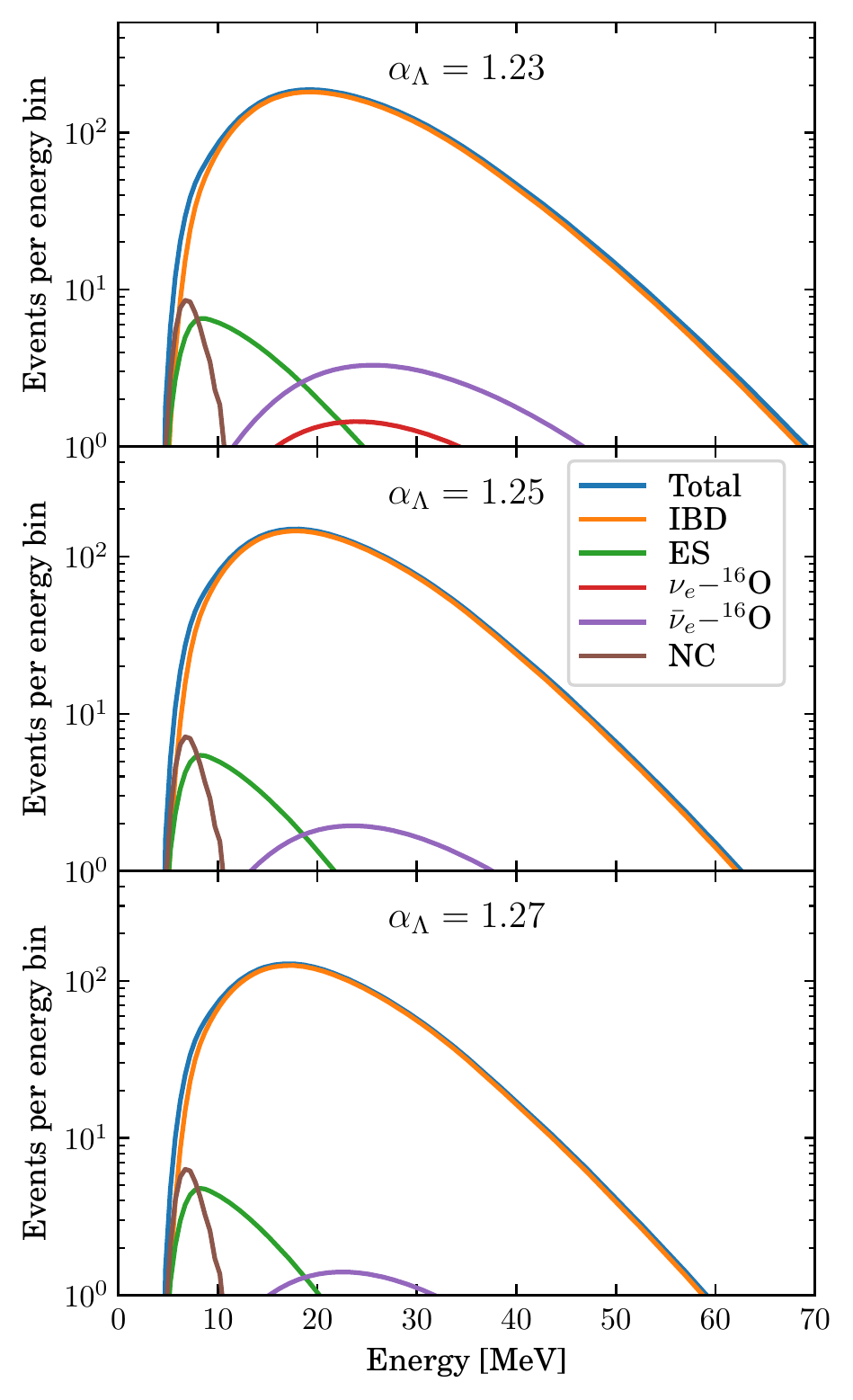}
\caption{Time-integrated event counts per energy bin out to 1s post-bounce for the $28\, \mathrm{M}_{\odot}$ progenitor from \cite{sukhbold:2016} for turbulence strength parameters $\alpha_{\Lambda} = 1.23$, 1.25, and 1.27.   This progenitor does not explode for $\alpha_{\Lambda} = 1.23$, but does for higher values of $\alpha_{\Lambda}$.  The neutrino energy spectrum is harder and has a higher number of counts for the non-exploding simulation. It is also clear from this figure how the different detection channels are sensitive to different neutrino energies.  Count rate is estimated for a $32 \,\mathrm{kton}$ water Cherenkov detector for a supernova event at $10 \,\mathrm{kpc}$, for energy bins of $0.5\,\mathrm{ MeV}$.  }
\label{fig:neut-spectra}
\end{figure}

We use \snowglobes~to model the detected neutrino signal  in a $32\,\mathrm{kton}$ water Cherenkov (Super Kamiokande-like) detector for a CCSN event at $10\,\mathrm{kpc}$, with time bins of $5\,\mathrm{ms}$ and energy bins of $0.5\,\mathrm{MeV}$.  We have used as input for \snowglobes~a pinched Fermi-Dirac neutrino spectrum reconstructed using parameterization of \cite{keil:2003}.  We use the ``smeared'' rates -- that is, the neutrino spectrum as a function of energy that will actually be observed in the detector --  for each of the relevant channels: inverse beta decay, electron scattering, charged current interactions with $^{16}\mathrm{O}$, and neutral current interactions for all flavors.  For details of how the event rates are calculated, we refer the reader to \cite{scholberg:2012} and references therin.    Table~\ref{tab:channels} outlines the relevant interaction channels, which are each sensitive to different neutrino flavors and energies.  The energy sensitivities and relative count rates of the different channels can also be seen in Figure~\ref{fig:neut-spectra}, which shows the time-integrated neutrino spectra to $1\,\mathrm{s}$ post-bounce for the $28\,\mathrm{M_\odot}$ progenitor from  \cite{sukhbold:2016}, for several values of the turbulence strength parameter $\alpha_{\Lambda}$.   The signal will be dominated by the inverse beta decay (IBD) channel, with other channels seeing $< 170$ counts in the first second post-bounce.     Additionally, the neutrino spectra are sensitive to whether or not the progenitor explodes.  This progenitor does not explode for $\alpha_{\Lambda} = 1.23$, but does explode for $\alpha_{\Lambda} = 1.25$ and $1.27$.  The neutrino spectrum is harder in all channels for the simulation that does not explode -- the observed neutrino mean energy is $24.4\,\mathrm{MeV}$ for $\alpha_{\Lambda} = 1.23$ compared to $22.5\,\mathrm{MeV}$ and $21.6\,\mathrm{MeV}$ for $\alpha_{\Lambda} = 1.25$ and $\alpha_{\Lambda} = 1.27$ respectively.  The simulation that does not explode also has a notably higher number of counts -- 9107 total counts for $\alpha_{\Lambda} = 1.23$, 6628 for $\alpha_{\Lambda} = 1.25$, and 5412 for $\alpha_{\Lambda} = 1.27$.    For the successful explosions, the $\nu_{e}-^{16}\mathrm{O}$ channel will have less than one count in the first second, while for the failed explosion it will have several counts.

Given the uncertainties in channel identification in real observing facilities and the large error bars in the count rates for many of these channels -- especially those with only a few counts -- we limit our discussion from here on to the total neutrino signal, that is the neutrino signal summed over all channels.  As shown in Figure~\ref{fig:neut-spectra}, all channels except for the IBD channel will have $<10$ counts in the first second.  Relying on so few counts to make predictions about the progenitor would lead to large uncertainties, given that that the statistical uncertainty on 10 counts would be $\sim \sqrt{N} \approx 32\%$.  With increased count rates in larger near future detectors such as Hyper Kamiokande, it may be possible to loosen this constraint in the future.

\
We ignore, for the purposes of this study, the effects of neutrino flavor mixing. There are a variety of flavor mixing phenomena that are expected to play a role in a CCSN signal: mixing in the CCSN mechanism due to collective oscillations \citep{duan2010}, fast oscillations \citep{sawyer:2005,sawyer:2009,martin:2020,morinaga:2020,johns:2020}, and MSW effects \citep{dighe:2000}, as well as mixing due to earth matter effects \citep{borriello:2012}.  It is to be expected that neutrino flavor mixing will drastically alter the observed neutrino signals and the expected correlations with progenitor and explosion properties.  However, given how little is known about these processes and how they occur in CCSNe, we cannot hope to completely  include them here and leave the sensitivities of these correlations to flavor mixing to future work.

\subsection{Gravitational waves\label{sec:methods-gw}}

We use the GW analysis detailed in \cite{morozova2018} to analyze the GW emission from the PNS.  This calculates the eigenfrequencies of the PNS, which are excited by convective downflows and other mechanisms to produce the observed GW emission.   Although GW signals are inherently produced by multidimensional effects, the dominant features in the GW signal are due to normal modes of the PNS -- particularly the quadrupole ($\ell = 2$) modes \citep{sotani:2016,sotani:2017,morozova2018,torres-forne:2018,torres-forne,torres-forne:2019,schneider:2019,sotani:2019,sotani:2019b}.

The analysis laid out in \cite{morozova2018} requires solving a system of equations including linearized general relativistic hydrodynamics.  A bisection method is used to find solutions that satisfy the imposed boundary condition that the pressure perturbation be zero at the PNS surface (assumed to be where $\rho = 10^{10} \mathrm{\, g \,cm}^{-3}$).  We direct readers to \cite{morozova2018} for details of how this calculation is done.  \cite{morozova2018} found that the difference between the Cowling and ``relaxed'' Cowling approximations was negligible at early times ($\lesssim 0.5\,\mathrm{s}$ post-bounce). We leave the validation of these two approaches for the future work and use the Cowling approximation here.

\begin{figure}[t!]
   \centering
   \includegraphics[width = 0.5\textwidth]{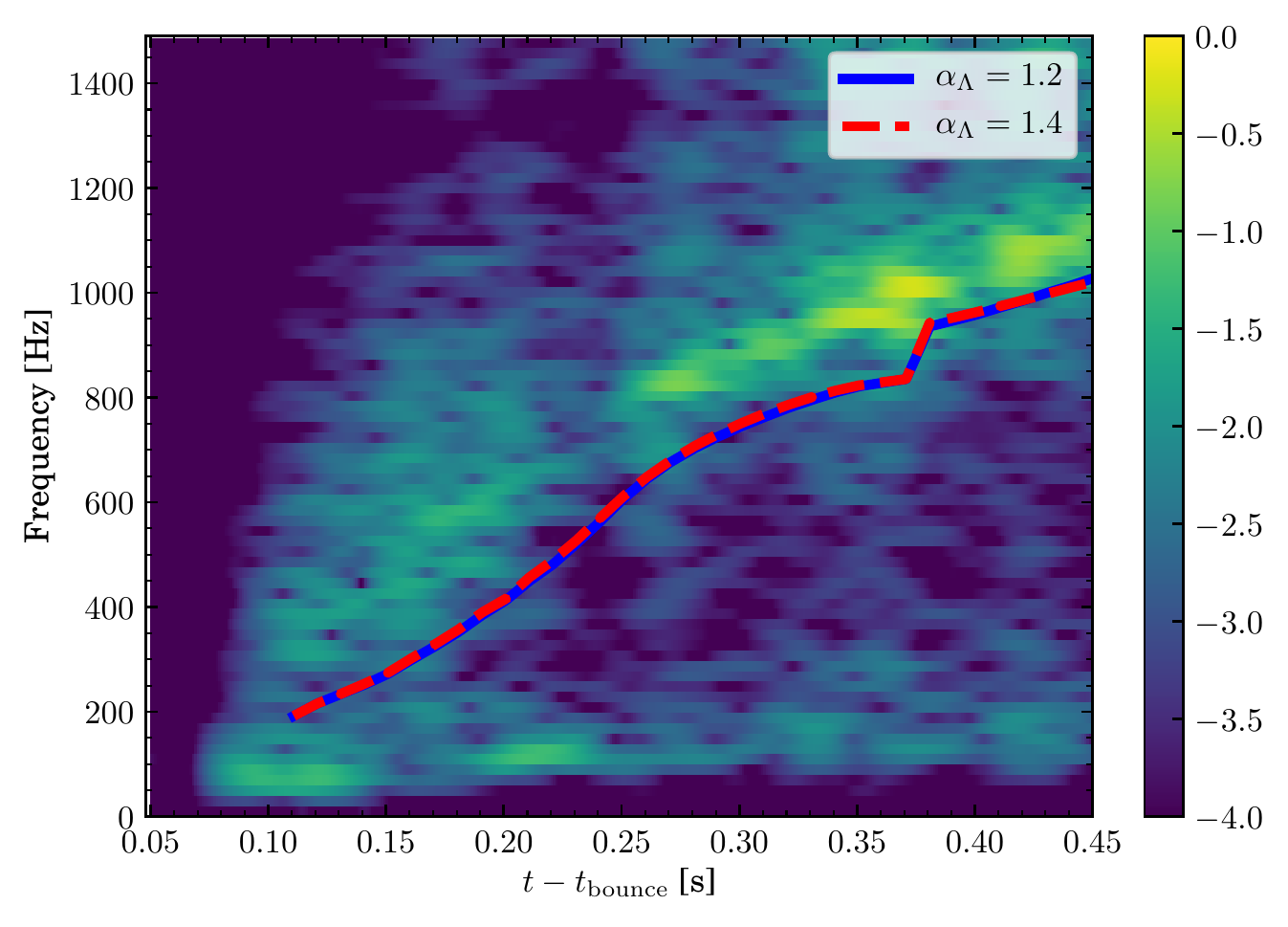} 
   \caption{GW spectrogram for 1D and 3D simulations for the $20\, \mathrm{M}_{\odot}$ progenitor of \cite{farmer:2016}.  The background 3D spectrogram is from simulations run in \cite{oconnor:2018}.  The blue line indicates the frequency evolution for the peak GW frequency found with the 1D GW code for the same progenitor, run with STIR using $\alpha_{\Lambda} = 1.2$, and the red line is the frequency evolution found in 1D with $\alpha_{\Lambda} = 1.4$.  Since there is little variation in the modes with $\alpha_{\Lambda}$, it is clear which mode is the dominant GW mode when comparing to the 3D spectrogram.}
   \label{fig:GWcomp}
\end{figure}

In our analysis, we are predominately interested in the behavior and evolution of the dominant frequency, which can be picked out by comparing to a GW spectrogram found in a 3D simulation (similar to the analysis done using 2D simulations in \cite{morozova2018}).  We do not concern ourselves with carefully identifying the specific mode type of this frequency band, but only care about tracking the evolution of the peak frequency, regardless of what mode is generating this frequency.  To determine which eigenfrequency corresponds with the peak emission, we compare to the fit formula of \cite{mueller:2013},
\begin{equation}
f_\mathrm{peak} = \frac{1}{2 \pi} \frac{G M_\mathrm{PNS}}{R_\mathrm{PNS}^{2} c} \sqrt{1.1 \frac{m_{N}}{\langle E \rangle_{\nu_{e}}}}\left(1 - \frac{G M_\mathrm{PNS}}{R_\mathrm{PNS} c^{2}}\right)^{2}~. \label{eq:peak}
\end{equation}
We determine which eigenfrequency lies closest to this expression, while also enforcing that the peak frequency cannot switch modes until the different between the mode frequencies is small.  Equation~\ref{eq:peak} was derived from the first $< 1\,\mathrm{s}$ post-bounce of 2D	 simulations.  Therefore, we do not trust it to properly approximate the peak frequency evolution at several seconds post bounce.  At later times, we assume that the peak frequency does not switch modes after $\sim0.5\,\mathrm{s}$ post-bounce, which seems to be an adequate assumption, since the modes diverge significantly at later times. We do not track the evolution of the mode prior to $100\,\mathrm{ms}$ post-bounce since, early post-bounce, there are many overlapping modes and eigenfrequencies, making it difficult to determine which frequency is the correct one to consider here.  The bounce signal may contain a wealth of information about the progenitor, rotation, equation of state, and numerous other quantities \citep{richers:2017,pajkos:2019}.  However, capturing the bounce signal is beyond what is feasible with 1D simulations and we do not concern ourselves with it here.  

We verify that we are able to pick out the peak frequency by comparing 1D STIR simulations of a $20\, \mathrm{M}_{\odot}$ progenitor to 3D simulations done in \cite{oconnor:2018}, as shown in Figure~\ref{fig:GWcomp}.  These simulations are for a $20\,\mathrm{M}_{\odot}$ progenitor created using the MESA stellar evolution code \citep{paxton2011,paxton2013,paxton2015,farmer:2016}, with the SFHo equation of state and M1 neutrino transport. 

The turbulence strength parameter will affect the GW emission since the turbulence will change the PNS structure by diffusing lepton fraction, including the temperature, and energy in the PNS. Additionally, turbulence will change the amount of fallback and accretion onto the PNS, thus changing its mass.  However, for early times ($<0.5 \,\mathrm{s}$) post-bounce for the simulations shown in Figure~\ref{fig:GWcomp}, with $\alpha_{\Lambda} = 1.2$, which does not explode, and $\alpha_{\Lambda} = 1.4$, there is negligible difference between the peak frequency evolution at early times and a $\sim11\%$ difference between the frequencies at about $2\,\mathrm{s}$ post-bounce.  Thus, we are able to pick out the dominant frequencies for all of the progenitors and turbulence strength parameters $\alpha_{\Lambda}$, assuming that this mode does not change dramatically between progenitors \citep{mueller:2013,yakunin:2015,morozova2018,torres-forne}.  The switch from one eigenfrequency mode to another is apparent in both the $\alpha_{\Lambda} = 1.2$ and $1.4$ simulations at about $0.40 \,\mathrm{s}$ post-bounce.  Since the eigenmodes are discrete, this switch in modes appears as sudden jump in the peak frequency. In multidimensional simulations, as is apparent in the spectrogram from the 3D simulation, the peak frequency is smooth and continuous and does not show a sudden jump.  This is likely because, as the peak frequency approaches the change in modes, both eigenmodes are simultaneously excited for some time, resulting in a smooth transition between frequencies.

It is worth noting that, while this analysis can reproduce the frequencies of the GW emission from the PNS, it fails to fully capture the GW emission seen in multidimensional simulations because it does not model the relative strength of each mode or the energy in each mode.  For example, the amplitude of the GW signal will be set by the convection, SASI, and other non-spherical hydrodynamic effects above the PNS \citep{pajkos:2019,radice:2019,andresen:2019,oconnor:2018b,kuroda:2016}.   There is some indication that the GW energy can be derived from the turbulent energy \citep{radice:2019}, but we leave such an analysis to future work.  We focus instead entirely on the evolution of the peak frequency of the GW emission without concerning ourselves with the relative strength of this frequency or the spectrum of different frequency bands.  Thus, the GW signals produced in this work, and by 1D models in general, cannot be used as exact predictions of GW signals from CCSNe.  However, knowledge of the peak frequency is sufficient for exploring the relationship between the GW emission and other physics of the supernova environment.

\subsection{Correlations\label{sec:methods-corr}}

We analyze the simulated neutrino and GW signals described above for correlations between observable quantities and fundamental physics parameters and properties of the progenitor stars.  Given the large sample of simulations we include here (600), we are uniquely situated to explore statistical correlations in a way that has been inaccessible to many previous studies due to small numbers or use of more approximate neutrino physics.  We take our inspiration for using correlation matrices to explore relationships between simulation properties from \cite{richers:2017}, which used correlation matrices to explore the relationship between equation of state parameters and the GW signal at bounce in rotating CCSNe.

We calculate the Pearson product-moment correlations coefficient using the Python SciPy package \citep{scipy}.   The Pearson coefficient $r$ is a measure of the linear correlation between two variables and is calculated by the covariance of the two variables of interest, divided by their standard deviations,
\begin{equation}
r = \frac{\sum_{i} \left(x_{i} - \bar{x}\right)\left(y_{i} - \bar{y}\right)} {\sqrt{\sum_{i} \left(x_{i} - \bar{x} \right)^{2} } \sqrt{ \sum_{i} \left(y_{i} - \bar{y} \right)^{2} } }
\end{equation}
for variables $x$ and $y$. A correlation coefficient value of $r\sim1$ indicates an extremely strong positive correlation, a correlation coefficient of $r\sim-1$ indicates an extremely strong negative correlation, and a coefficient value of $r\sim0$ indicates that there is no correlation between the variables. For the purposes of this work, we consider a strong correlation to be $|r| \gtrsim 0.5$, a moderate correlation to be $0.3 \lesssim |r| \lesssim 0.5$, and $ |r| \lesssim 0.3$ to be a poor correlation.   Note that the $r$ value does not indicate the \textit{slope} of relationship between the data, but is a measure of the strength of the correlation.  

This method of using the Pearson $r$ value will not distinguish between ``noisy'' linear data and data related in non-linear ways - for example, a noisy linear relationship can have the same $r$ value as a tight, quadratic relationship.  It is unlikely that all, or even most, quantities of interest will be linearly correlated in such a complex environment as a CCSNe.  Considering linear correlations will provide a significant first step and, as we will see, be sufficient to uncover insightful and significant relationships between observable quantities and progenitor and explosion properties \citep[cf.,][]{torres-forne}.  More complex analysis to determine nonlinear correlations is left for future work. 

In conducting this analysis, we consider  numerous variables and parameterizations of the progenitor, explosion, and observable properties.   There are countless quantities that could be considered here.  We have explored numerous parameters of the neutrino signal, progenitor, and explosion.  Many of these parameters were found to have little relationship with the physics that we are interested in exploring here, such as the radius of the star at the time of collapse, or are intimately related with quantities that we do consider.  For example, we will use the core compactness $\xi_{2.5}$ \citep{oconnor:2011}, 
\begin{equation}
\xi_{2.5} = \left.\frac{2.5\, \mathrm{M}_{\odot} /\mathrm{M}_{\odot}}{R(M_\mathrm{bary} = 2.5\, \mathrm{M}_{\odot})/1000\,\mathrm{ km}} \right|_{t = t_\mathrm{collapse}} ,
\label{eq:comp}
\end{equation}
as a measure of the structure of the progenitor core at the time of collapse.  Similar parameters, such as the parameters defined by \cite{ertl:2016}, measure essentially the same thing and would have very similar relationships with observable quantities.  Other parameters are known to be unrelated from previous works; for example, the ZAMS mass has little relationship with the explosion and physics in the core \citep{oconnor:2011,ugliano:2012,ertl:2016,sukhbold:2016,couch:2020}  and the peak luminosity is largely insensitive to the ZAMS mass \citep{liebendorfer:2004,oconnor:2013}.   For sake of clarity and brevity, we will only show variables here that showed strong (anti-)correlations with observables.   For a list of parameters considered, please see Appendix~\ref{sec:parameters}.

\section{Results\label{sec:results}}

\subsection{Distinguishing failed versus successful events\label{sec:expld}}

Perhaps the simplest ``observable'' feature from the death of a massive star is determining whether or not the core collapse results in a successful supernova event or a failed supernova and collapse to BH.  The critical difference between these possibilities is of course the occurrence of an electromagnetic transient consistent with a CCSN.  However, there may be cases where such a transient is not visible if it occurs toward the galactic center.  Also, given that the electromagnetic transient occurs $\sim$hours-days after the GW and neutrino emission, one might hope to determine the outcome of the core collapse without waiting for shock breakout.

\begin{figure}
\includegraphics[width = 0.47\textwidth]{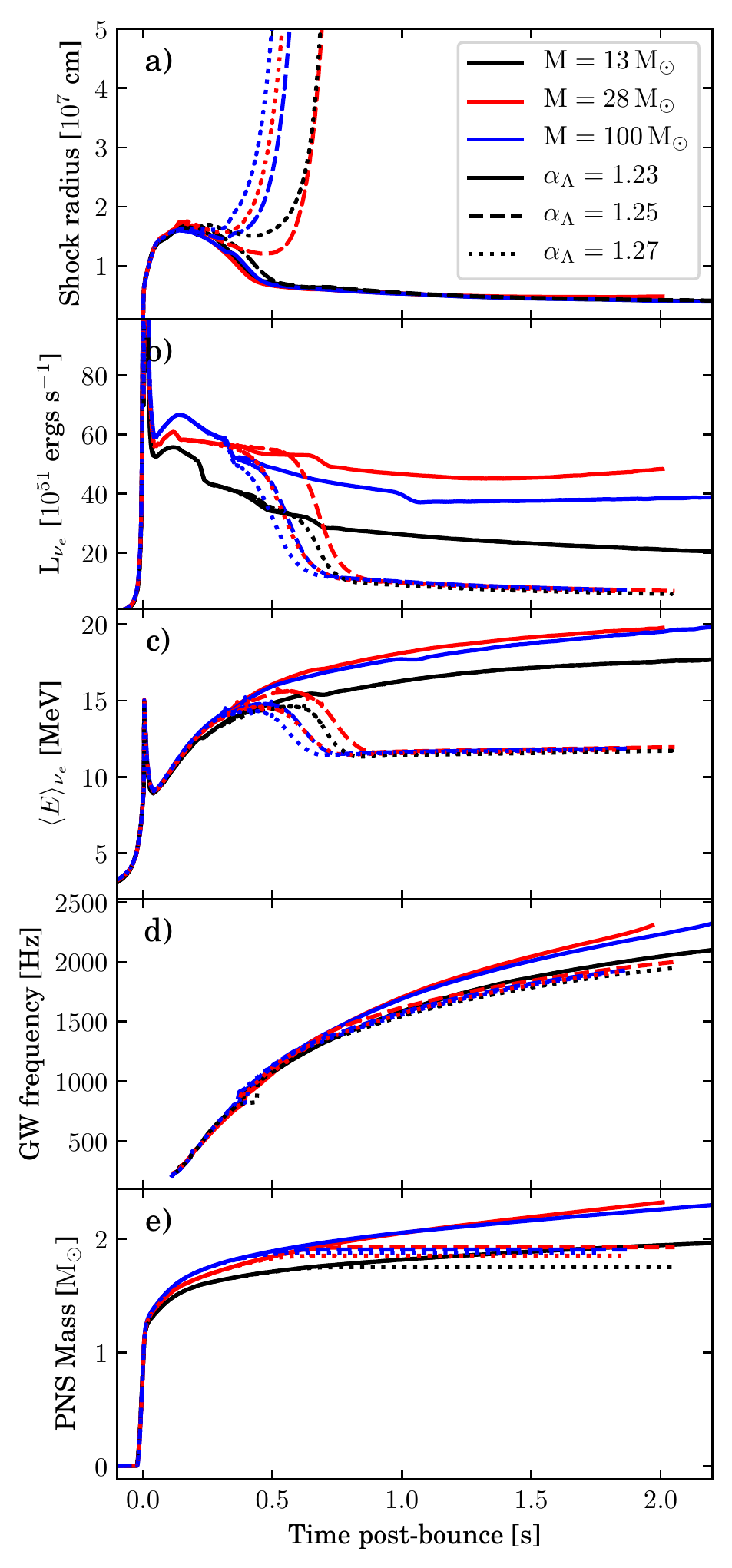}
\caption{Shock radius, PNS mass, and multi-messenger signals versus time post-bounce for various progenitors and turbulence strength parameters.  Panel a shows the shock radius, panel b shows electron neutrino luminosity, panel c shows the electron neutrino mean energy, panel d shows the dominant GW frequency, and panel e shows the PNS (baryonic) mass..  The black lines are for $13\, \mathrm{M}_{\odot}$ progenitor, the red for $28 \,\mathrm{M}_{\odot}$, and blue for a $100 \,\mathrm{M}_{\odot}$ progenitor.  The solid lines are a turbulence strength parameter of $\alpha_{\Lambda} = 1.23$, the dashed lines show $\alpha_{\Lambda}=1.25$, and the dotted lines $\alpha_{\Lambda} = 1.27$.  The $13\, \mathrm{M}_{\odot}$ progenitor has a critical $\alpha_{\Lambda}$ value of 1.27 at which it begins to explode.  The $28 \,\mathrm{M}_{\odot}$ and $100 \,\mathrm{M}_{\odot}$ progenitors have a critical $\alpha_{\Lambda}$ value of 1.25. It may be possible to tell when explosion sets in by the sharp drop off in the neutrino luminosity and mean energy, which corresponds with the diverging GW signals for each progenitor.  Note that, although not plotted here, there will be GW emission prior to 200 ms post-bounce.  }
\label{fig:explvfail}
\end{figure}

There are a few key observable differences between the neutrino and GW emission from successful and failed events.  Figure~\ref{fig:explvfail} shows the shock radius, neutrino luminosity, neutrino mean energy, peak GW frequency, and baryonic PNS mass versus time post-bounce for failed and successful CCSN events for several progenitor masses.  These progenitors were chosen because they span the range of ZAMS mass (13, 28, and 100 $\mathrm{M}_\odot$) and core compactness (0.058, 0.281, and 0.244 respectively)  and have a ``critical'' $\alpha_{\Lambda}$ value between 1.23 and 1.27.   For a given progenitor, there is a critical $\alpha_{\Lambda}$ value at which it starts to explode, the value of which depends sensitively on the progenitor \citep{couch:2020}.  

The most obvious distinction between successful and failed events will be an abrupt ``stop'' in the neutrino and GW signal.  The GW emission should stop abruptly, similar to what is seen in simulations of BH formation for binary neutron star mergers (for example see \cite{radice:2017}). When the cooling PNS collapses to a BH, neutrino emission should end quite promptly as well. Thus, a sharp ending of both the neutrino and GW emission will be a clear indicator of a failed explosion.

In our 1D models, the neutrino luminosity and mean energy drop off at the time that explosion sets in, as evident in comparing the neutrino emission to the shock radius.  This is due to the fact that accretion onto the PNS stops once the explosions sets in and can no longer serve as a source of neutrinos, leaving only the cooling PNS as a neutrino source.  One could imagine using the observation of such a drop off as a marker of the time of the onset of explosion.  However, this would depend on being able to distinguish a drop off in neutrino emission due to the onset of explosion from a drop off due to the structure of the progenitor star, e.g.,~the accretion of the Si/O interface through the shock \citep{seadrow:2018}, although there are some indications that the accretion of the Si/O interface is tied to the explodability  \citep{fryer:1999,murphy:2008,summa:2016, vartanyan:2018} . For example, the $13\, \mathrm{M}_{\odot}$ simulations in Figure~\ref{fig:explvfail} don't explode for $\alpha_{\Lambda} = 1.23$ and $\alpha_{\Lambda} = 1.25$.  Yet, there is still a ``drop'' in the neutrino luminosities at $\sim0.25\,\mathrm{s}$ post-bounce due to the structure of the progenitor star.  A good diagnostic to distinguish between the onset of explosion and accretion of the Si/O shell interface seems to lie in the neutrino average energies.  The average energies, shown in panel c of Figure~\ref{fig:explvfail}, do not show the same dramatic drop as the neutrino luminosities at the accretion of composition interfaces, only at the onset of explosion.  Thus, observation of a sharp drop off in both the neutrino luminosity and mean energy could be indicative of the onset of explosion. This is in agreement with the results of \cite{sumiyoshi:2006} and \cite{sumiyoshi:2008}, where they found that the increased accretion in failed CCSN events leads to higher neutrino luminosities and average energies after the neutronization burst relative to a successful explosion.

Further work must be done to explore this phenomenon thoroughly in multidimensional models.  The drop off in accretion at the onset of explosion is stronger in 1D simulations than in 2D or 3D, since multidimensional models allow for continued accretion onto the PNS surface after the explosion begins, but this feature appears to still present in most multidimensional simulations \citep{oconnor:2018b,mueller:2019a,burrows:2019}.  It remains unexplored how these observable features depend on the line of sight, and perhaps more importantly, whether the line of sight is aligned or misaligned with a downflow of accreted material.

There is also a divergence in the evolution of the peak GW frequency, shown in panel d of Figure~\ref{fig:explvfail}, at the onset of explosion.  As a given progenitor reaches its critical $\alpha_{\Lambda}$ value, at which it starts exploding, the peak GW frequency diverges at the time of the onset of explosion and asymptotes to lower frequencies than in the case of a failed supernova event.  This is much less dramatic effect than that seen in the neutrino signal. It will likely be difficult to distinguish this effect given the differences in peak GW behavior between progenitors -- the failed event for the $13\, \mathrm{M}_{\odot}$ progenitor is difficult to distinguish from the successful explosions of the $28$ and $100\, \mathrm{M}_{\odot}$ progenitors.  The divergence of the GW signals is related to the PNS structure and evolution.  Panel~e in Figure~\ref{fig:explvfail} shows the evolution of the baryonic PNS mass for these simulations. The same divergence occurs in the PNS masses with the onset of explosion. This is due to the fact that, as noted above, in 1D accretion shuts off at the onset explosion, thereby halting the growth of the PNS mass.  This steeper rise in the GW frequency for failed CCSNe was also seen by \cite{sotani:2019}.

The success, or failure, of a nearby CCSN event should be imprinted on the neutrino and GW signals long before EM emission is visible.  Failed events result in a sharp end to the neutrino and GW emission in the first few seconds.  Successful events have a drop in the neutrino counts and mean energy at the time that explosion sets in, but have continued neutrino and GW emission.  From now on, we consider ``successful'' and ``failed'' CCSNe as separate populations.  As we will see, the relationship between observable parameters and information about the progenitor are very different in the two cases due to the different physics driving the creation of neutrinos and GW signals in these environments.

\subsection{Neutrino signals\label{sec:neutrino}}

We consider the neutrino signal that would be observed in a Super Kamiokande-like $32\, \mathrm{kton}$ water Cherenkov detector for a CCSN event at $10\,\mathrm{kpc}$, as simulated using the \snowglobes~code \citep{scholberg:2012}.  Using \snowglobes, we are able to obtain the time-evolution of the observed neutrino spectra in the various channels, as listed in Table~\ref{tab:channels}.

\begin{figure*}
\centering
  \includegraphics[width = \textwidth]{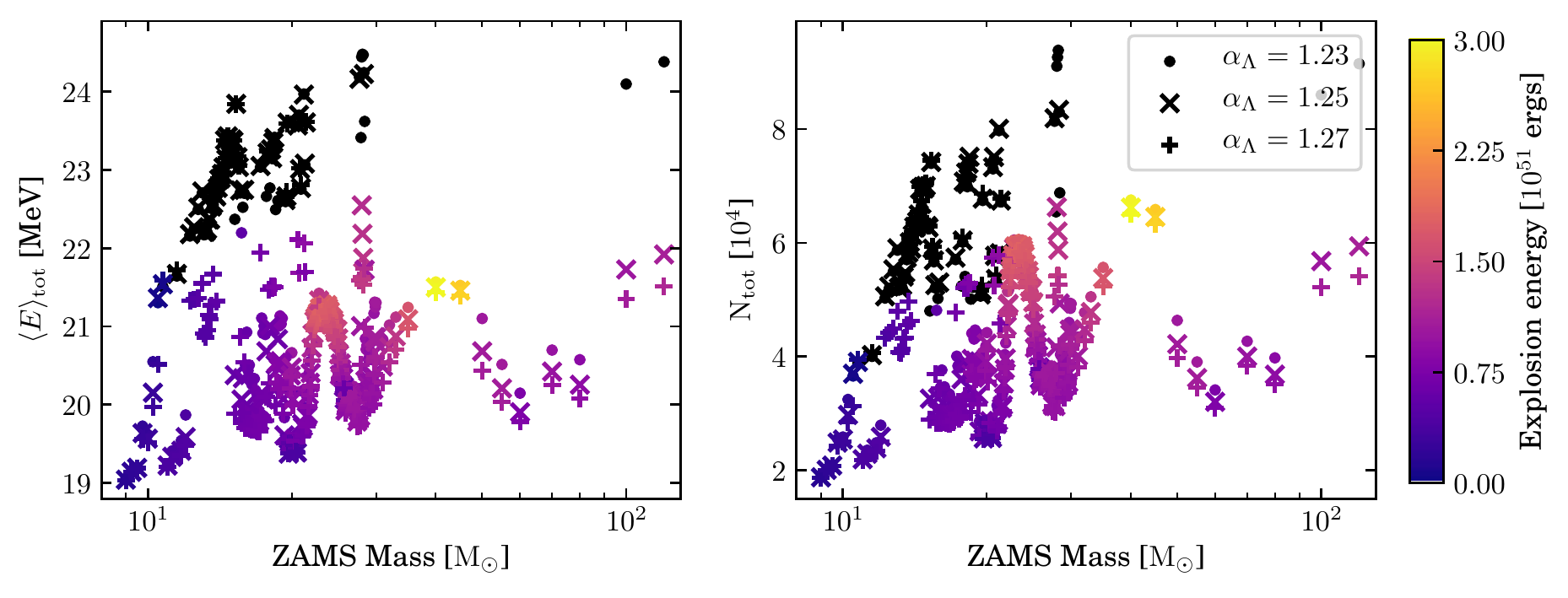}
  \caption{Left panel shows the time-integrated neutrino mean energy, $\langle E \rangle_\mathrm{tot}$, and the right panel shows the time- and energy-integrated counts, $\mathrm{N}_\mathrm{tot}$, to $1\,\mathrm{s}$ post-bounce and summed over all channels, versus ZAMS mass.  Energies and count rates were calculated assuming a $32\, \mathrm{kton}$ water Cherenkov detector for a CCSN event at $10 \,\mathrm{kpc}$.   The color scales with the explosion energy and black indicates a simulation that did not explode, including those that ran to $5\,\mathrm{s}$ post-bounce without collapsing to a BH.  The symbols indicate the turbulence strength parameter $\alpha_{\Lambda}$.  Neither average energy nor neutrino counts show any simple relationship with the ZAMS mass.  There is a somewhat progenitor-dependent dividing line in average energy that separates successful explosions from those that fail. However, explosion energy appears to be more strongly related to the neutrino counts.}
  \label{fig:neut-parameters}
\end{figure*}

In order to simplify comparison between simulations, we do not consider here the time evolution of the neutrino emission.  There are two parameters that we consider: the mean energy $\langle E_\mathrm{tot} \rangle$ of the time-integrated detected neutrinos and the time- and energy-integrated number of neutrinos detected, $\mathrm{N}_\mathrm{tot}$, or counts, in all channels, integrated to $1\,\mathrm{s}$ post-bounce. We note that for the former, due to the energy dependence of the neutrino cross sections in water, the mean detected energy is not the same as the mean energy of the emitted spectrum (i.e. the values plotted in Figure~\ref{fig:explvfail}).  Although we have chosen to integrate out to $1\,\mathrm{s}$ post-bounce, one should note that, for successful CCSN events, the neutrino signal may be detectable for 10s or 100s of seconds, and the late time signal will contain information about the PNS mass and structure \citep{roberts:2012,suwa:2019}.  There are many indications though that significant information about the progenitor star and explosion can be inferred from the first second or less of the neutrino signal (e.g.~see \cite{horiuchi:2017}). We find that there were no significant correlations between basic measures of the time evolution, such as timescales, slopes, and fits of the neutrino quantities.  Using time-integrated quantities has the added benefit of decreasing potential observational uncertainties - the higher the number of counts being considered, the smaller the relative error.  This will allow for a more reliable relationship to be established between observable quantities and the underlying progenitor structure and explosion.

\begin{figure}[b]
\centering
\includegraphics[width = 0.5 \textwidth]{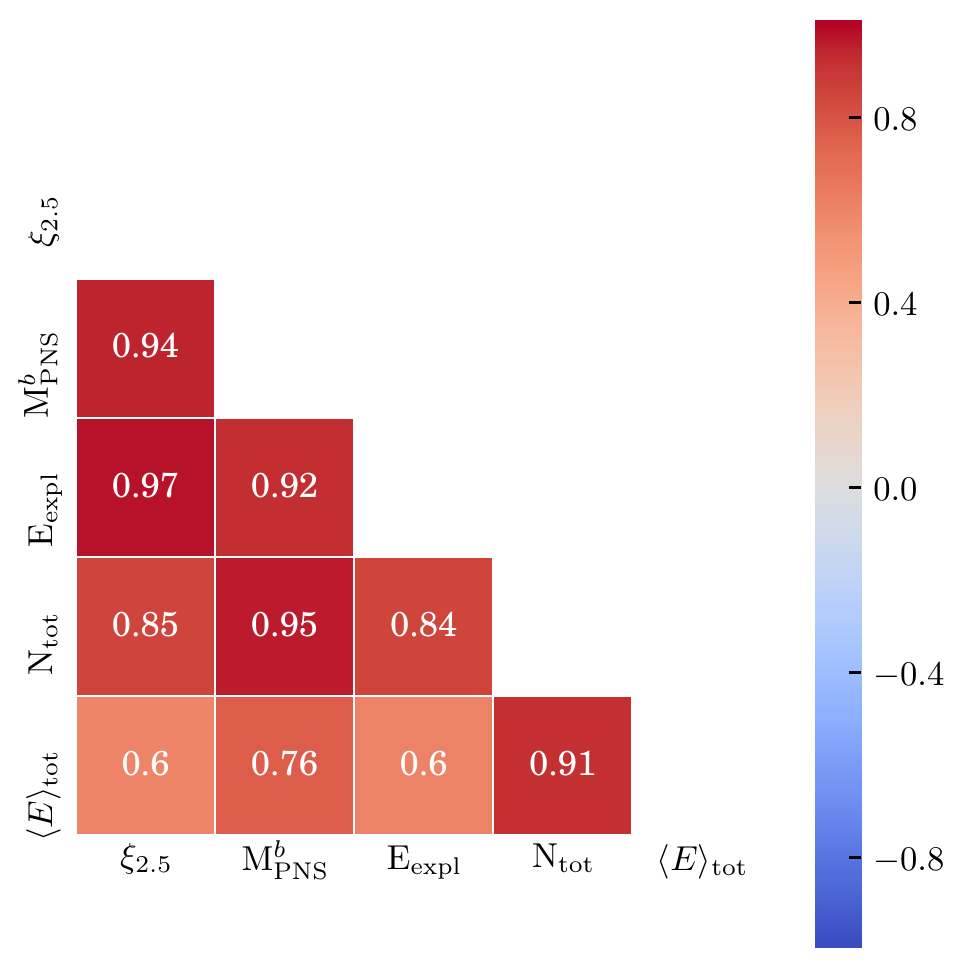}
\caption{Correlation matrix for observable neutrino quantities and underlying properties of the progenitor and explosion for simulations that successfully explode.  The quantities considered are:  core compactness ($\xi_{2.5}$), baryonic PNS mass ($\mathrm{M}^{b}_\mathrm{PNS}$) at 1s post-bounce, explosion energy ($\mathrm{E}_\mathrm{expl}$),  time- and energy-integrated neutrino counts ($\mathrm{N_{tot}}$) in all channels, and the time-integrated average neutrino energy in all channels ($\langle E\rangle_\mathrm{tot}$).  The lower-left half shows the correlation strength, ranging from -1 to 1, as indicated by the color and number in each of the squares. } 
\label{fig:neut_ns}
\end{figure}

We can now ask whether we can distinguish features of the supernova progenitor using these time-integrated neutrino quantities.  Figure~\ref{fig:neut-parameters} shows the mean energy and counts in all channels for progenitors with ZAMS mass between $9\,\mathrm{ M}_{\odot}$ and $120 \,\mathrm{M}_{\odot}$ and turbulence strength parameter $\alpha_{\Lambda}$ between 1.23 and 1.27, with the explosion energy indicated by the color.  As has been seen with other CCSN observables, such as the explodability shown in Figure~\ref{fig:stir}, the neutrino mean energy and counts do not vary monotonically with ZAMS mass.  The neutrino mean energy, similar to the explodability, shows a behavior with ZAMS mass similar to the compactness parameter $\xi_{2.5}$ \citep{oconnor:2011}, as defined in Equation~\ref{eq:comp}.
 This is perhaps unsurprising, as the neutrino emission comes directly from the core of the CCSN event and is not significantly impacted by the ZAMS mass or outer envelope.

The average detected neutrino energy, shown in the left panel of Figure~\ref{fig:neut-parameters}, and neutrino counts, in the right panel, show a general decreasing trend with increasing turbulence strength for a given progenitor.  This is consistent with previous work done that shows decreasing neutrino average energies with turbulence strength in the STIR model \citep{couch:2020}.  The rate, however, that the mean energy and counts decrease with turbulence strength varies.  This is related to the explodability and critical \alphL value of each particular progenitor.  The neutrino mean energy and counts drop dramatically at the critical \alphL value for a given progenitor - that is, they drop dramatically in simulations that explode successfully compared to simulations that fail. This can be seen most clearly for the $100$ and $120\, \mathrm{M}_\odot$ progenitors, which have a critical \alphL value of 1.25 at which they begin to explode.  However, this is not a reliable means to distinguish between a successful and failed events since the ``dividing line'' in mean energy and counts between failed and successful CCSNe varies dramatically between progenitors.

We can begin to explore the deeper connections between these two detectable neutrino parameters -- mean energy and counts -- with underlying properties of the progenitors, explosion mechanism, and remnant compact object.    Figure~\ref{fig:neut_ns} shows the correlation strength for simulations that explode successfully.  We compare here the behavior of the neutrino emission with the core compactness ($\xi_{2.5}$), the PNS mass ($\mathrm{M}^{b}_{\mathrm{PNS}}$) at 1 s post-bounce, and the explosion energy ($\mathrm{E}_\mathrm{expl}$).  The PNS mass considered throughout this work is the baryonic mass.  The core compactness ($\xi_{2.5}$) is strongly correlated with the number of neutrino counts in all channels ($\mathrm{N_{tot}}$), as was also seen by \cite{horiuchi:2017}.  The core compactness is related to the total gravitational energy released in the collapse and thus, the total neutrino energy released \citep{oconnor:2013}.   The PNS mass at $1\,\mathrm{s}$ post-bounce ($\mathrm{M}^{b}_\mathrm{PNS}$) is also strongly correlated with the number of neutrino counts in all detection channels, as shown in the second column of Figure~\ref{fig:neut_ns}.   

Explosion energy is also strongly correlated with the number of neutrino counts.  The explosion energy is highly sensitive to the amount of neutrino heating in the gain region behind the shock - and the heating rate scales with the neutrino energies and counts.  The count rate in the detector is similarly sensitive to these quantities, given that the detector will be more sensitive to higher energy neutrinos.  Thus, the count rate in the detector may give significant insight into the neutrino heating occurring in the CCSN environment.  The explosion energy is less strongly correlated with the neutrino mean energy, $\langle E \rangle_\mathrm{tot}$, than with the neutrino counts.  This is also somewhat apparent in Figure~\ref{fig:neut-parameters}.  The ``peak'' in the explosion energy around ZAMS mass of $40\,\mathrm{M}_{\odot}$ is also a peak in the number of counts, for simulations that do explode successfully, but is less pronounced of a peak in the neutrino mean energy, $\langle E \rangle_\mathrm{tot}$.

\begin{figure}[b]
\centering
\includegraphics[width = 0.5\textwidth]{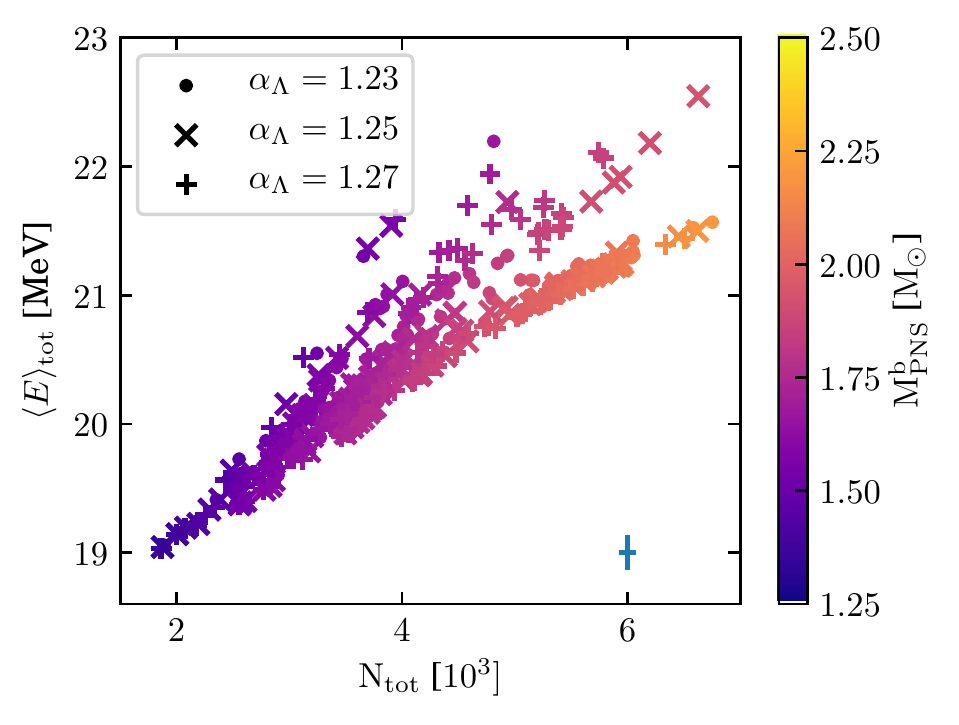}
\caption{The plane of  neutrino mean energy ($\langle E \rangle_\mathrm{tot}$) of the time-integrated observed neutrino spectrum versus time- and energy-integrated neutrino counts ($\mathrm{N_{tot}}$), in all channels out to $1\,\mathrm{s}$ post-bounce,  for simulations that explode successfully.  Count rates are simulated for a $32\,\mathrm{kton}$ water Cherenkov detector for a CCSN event at $10\,\mathrm{ kpc}$.  The PNS baryonic mass $\mathrm{M}^{b}_{\mathrm{PNS}}$ at 1s post-bounce is indicated by the color bar. Symbols indicate the turbulence strength parameter $\alpha_{\Lambda}$ used in the simulation.   The blue shows an example of the observational error bars for measuring the total number of counts and mean energy at $10\,\mathrm{kpc}$. }
\label{fig:neut-nsmass}
\end{figure}

 Given the strong correlations between the neutrino counts and quantities that cannot be directly observed, such as the PNS mass, it may then be possible to obtain these progenitor and explosion properties with careful observations of the neutrino emission.  Figure~\ref{fig:neut-nsmass} shows the plane of neutrino counts and average neutrino energy at 1 second after bounce for all of our simulations that explode successfully, with the PNS mass indicated by the color.  There is a clear trend in PNS mass with neutrino counts and mean energy, in concordance with the high values of the correlation parameter found earlier.  
 
 For a \textit{fixed} number of neutrino counts, there is a counterintuitive trend that increasing mean energy corresponds with decreasing PNS mass, contrary to the positive correlation found in these quantities in Figure~\ref{fig:neut_ns}.  This can be understood as being due to competing effects.  The number of neutrino counts is integrated over the first second post-bounce and, for a given, fixed number of neutrino counts, there are several ways to generate that number -- a high luminosity over a shorter period, or lower luminosity over the full $1\,\mathrm{s}$.  Simulations that result in a higher PNS mass tend to have a higher compactness and explode more quickly, thus having a higher luminosity but over a shorter period until the explosion sets in ($t_\mathrm{expl} < 1\,\mathrm{s}$).  The mean energy, however, is initially quite similar for all of these simulations, until it drops at the onset of explosion.  Thus the mean energy integrated over $1\,\mathrm{s}$ post-bounce will be lower for simulations that explode more quickly.  For a fixed number of neutrino counts, a higher PNS mass will correspond to a lower neutrino mean energy.  It is clear that the CCSN environment is not set or discernible using a single parameter alone, be it core compactness, PNS mass, etc.  There are competing effects that lead to a great deal of complexity in these simulations.
 
 The lack of scatter despite using several values of the turbulence strength parameter $\alpha_{\Lambda}$ indicates that this relationship is relatively insensitive to the assumptions and uncertainties in our underlying 1D supernova model.  
 Significantly, this is a measure of the PNS mass at $1\,\mathrm{s}$ post-bounce \textit{as the explosion is occurring} and \textit{before fallback accretion}. Such a measurement would provide a means to ``watch'' the evolution of the PNS and help determine the role of fallback accretion in the CCSN environment.  A similar process can be applied to determining other progenitor and explosion parameters, such as the core compactness and explosion energy.  
 
\begin{table*}[t]
   \centering
   \topcaption{Linear fits to observable neutrino parameters for successfully exploding CCSN events for the core compactness, $\xi_{2.5}$, PNS mass at $1\,\mathrm{s}$ post-bounce, $\mathrm{M}^{b}_\mathrm{PNS}$, and explosion energy, $\mathrm{E}_\mathrm{expl}$.  These coefficients are for functional fit of the form $ y= \mathrm{C_{N_{tot}}} \mathrm{N_{tot}}+\mathrm{C}_{\langle E \rangle_\mathrm{tot}} \langle E \rangle_\mathrm{tot}+ \mathrm{D}$, with the mean energy, $\langle E\rangle_\mathrm{tot}$, measured in MeV.  $R^{2}$ is the correlation coefficient between the data and the fit, and measures the quality of the fit. Note that these fits assume a distance of $10\,\mathrm{kpc}$ and thus the number of counts will have to be scaled accordingly for CCSN events at other distances.  } 
   \begin{tabular}{@{} l c c c c@{}} 
      \toprule
      Parameter &  C$_\mathrm{N_{tot}}$ &C$_{\langle E \rangle_\mathrm{tot}}$ & D & $R^{2}$ \\
      \midrule
     $\xi_{2.5}$ [unitless] &    $(8.662 \pm 0.255) \times 10 ^{-5}$ & --- & $(-1.272 \pm 0.107 ) \times 10^{-1}$ & 0.714 \\
        &   --- &  $(1.011 \pm 0.063) \times 10^{-1}$ & $-1.845 \pm 0.129$  & 0.361 \\
     $\mathrm{M}^{b}_\mathrm{PNS}$ [$\mathrm{M}_{\odot}$] &   $(1.643\pm 0.024) \times 10^{-4} $ & ---&  $1.093 \pm 0.010$ & 0.911  \\
      & --- &  $(2.156 \pm 0.009) \times 10^{-1}$ & $-2.656 \pm 0.175$ & 0.582 \\
     $\mathrm{E}_\mathrm{expl}$ [10$^{51}$ erg] &   $(3.255 \pm 0.096) \times 10^{-4}$ & --- & $(-4.380 \pm 0.405) \times 10^{-1}$  & 0.712\\
      & --- & $(3.811 \pm 0.236) \times 10^{-1}$ & $-6.921 \pm 0.483$ & 0.362 \\
       \bottomrule
   \end{tabular}
   \label{tab:expl_neut_fits}
\end{table*}

 In Table~\ref{tab:expl_neut_fits}, we provide linear fits between the observable neutrino parameters and the core compactness, PNS mass, and explosion energy for successfully exploding CCSN events, of a functional form
 \begin{equation}
 y= \mathrm{C_{N_{tot}}} \mathrm{N_{tot}}+\mathrm{C}_{\langle E \rangle_\mathrm{tot}} \langle E \rangle_\mathrm{tot}+ \mathrm{D}~,
 \end{equation}
where $\mathrm{C_{N_{tot}}}$ is the coefficient to the number of neutrino counts $\mathrm{N_{tot}}$, which can be found in Table~\ref{tab:expl_neut_fits} and $\mathrm{C}_{\langle E \rangle_\mathrm{tot}}$ is the coefficient to the neutrino mean energy $\langle E \rangle_\mathrm{tot}$.
     These fits can be used to estimate progenitor and explosion properties from a galactic supernova event and require just the first second of neutrino detections.  These fits assume a distance of $10\,\mathrm{kpc}$.   We have not included fits using both the number of neutrino counts, $\mathrm{N_{tot}}$, and mean energy $\langle E \rangle_\mathrm{tot}$ because, due to the stronger correlation with counts than mean energy, the coefficients of the best fit for the mean energy being essentially zero and the fit relies on the counts alone. 
 
Such a detection and early parameter estimation will provide insight into the progenitor star without requiring pre-explosion imaging.  Perhaps most significantly, measurement of the core compactness, $\xi_{2.5}$, will provide insight into the structure and core of the progenitor star that can be compared with what is found with pre-explosion imaging.  An additional measurement constraining the core structure will put significant constraints on stellar evolution models and the evolutionary tracks of CCSN progenitors.

\begin{figure*}[t]
\centering
\includegraphics[width =\textwidth]{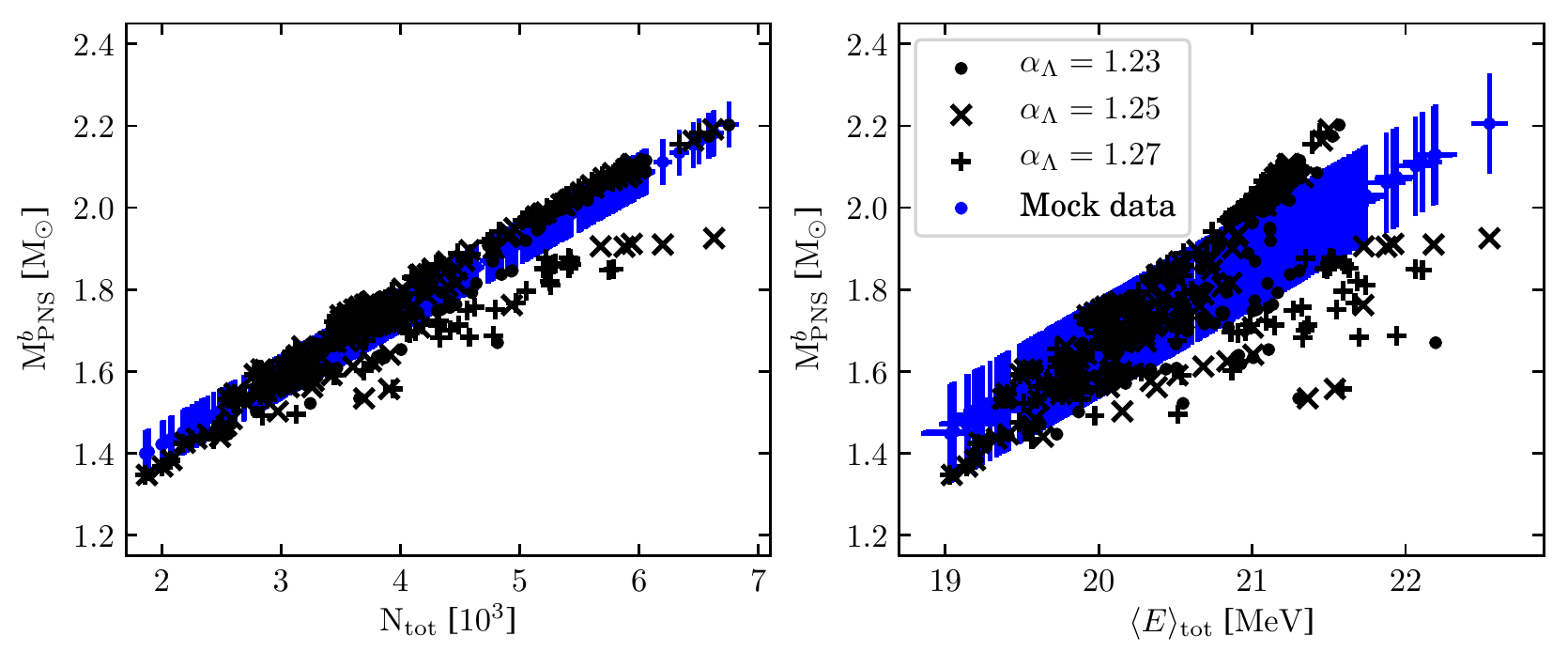}
\caption{Left panel shows the PNS baryonic mass ($\mathrm{M^{b}_{PNS}}$) versus time- and energy- integrated neutrino counts in all channels out to 1s post-bounce ($\mathrm{N_{tot}}$).  The right panel shows the PNS mass versus  neutrino mean energy of the time-integrated observed neutrino spectrum in all channels out to 1s post-bounce ($\langle E \rangle_\mathrm{tot}$).  Black points are the simulation results for $\alpha_{\Lambda} = 1.23 - 1.27$.  Blue points show the PNS mass found using the fitted relationship from Table~\ref{tab:expl_neut_fits}.  Fitted points include the statistical error in the measurement and standard deviation of the residuals.  It is clear that, for a distance of $10\,\mathrm{kpc}$ at least, it is advantageous to determine the PNS mass from the number of counts, rather than the mean energy. However, using the number of counts in this determination depends on having an accurate measurement of the distance.}
\label{fig:mockdata_both}
\end{figure*}

 Figure~\ref{fig:neut-nsmass} also includes an example of the purely statistical error bars for the measurement of the number of neutrino counts and mean energy (but assuming no error on the individual neutrino energy measurements, which varies by detector)  and Figure~\ref{fig:mockdata_both} shows the PNS mass found using the fitted relationship in Table~\ref{tab:expl_neut_fits}, including the errors associated with the fitting and the statistical errors associated with the measurement of the total counts or mean energy.  The scatter in the simulation points (black) in these plots is indicative of the range of the time for onset of explosion among the different progenitors.  Those close to the linear clustering have explosion times $< 1\,\mathrm{s}$ post-bounce, while those scattered to higher neutrino count and mean energy values have progressively longer explosion times.  Thus, by measuring the number of neutrino counts or mean energy only out to $1\,\mathrm{s}$ post-bounce, we are probing a smaller and smaller fraction of the total neutrino energy to be released, which in turn is a probe of the gravitational binding energy of the PNS.  While we could instead have measured the neutrino emission out to the time of explosion, or to the end of the simulation, these are points of time that may be difficult to discern when observing a real CCSN event (or in the case of the end of the simulation -- a completely arbitrary length of time).  Measuring the neutrino counts and average energies lets us compare between simulations using a timescale that can easily be applied to a real CCSN detection. 
 
 From Figure~\ref{fig:mockdata_both}, it is clear that, at a distance of $10\,\mathrm{kpc}$ at least, the error is somewhat reduced when using the number of neutrino counts rather than the mean energy.  This is reflected in the larger $R^{2}$ value of the fit (0.361 for mean energy compared to 0.714 for number of counts) and the strength of the correlations (0.6 for mean energy compared to 0.85 for number of counts). Thus one must take care when considering the fits, such as those shown in Table~\ref{tab:expl_neut_fits}, to also consider what the observational errors may be and use the most beneficial relationship.  The fits provided here assume a distance of $10\,\mathrm{kpc}$, so CCSN event at a further distance with fewer counts or an event with significant uncertainties in the distance estimation, will not provide as reliable of a determination of the core compactness as indicated by Figure~\ref{fig:neut-nsmass}, which may make it more precise to use mean energy.

\begin{figure}[b]
\centering
\includegraphics[width = 0.5 \textwidth]{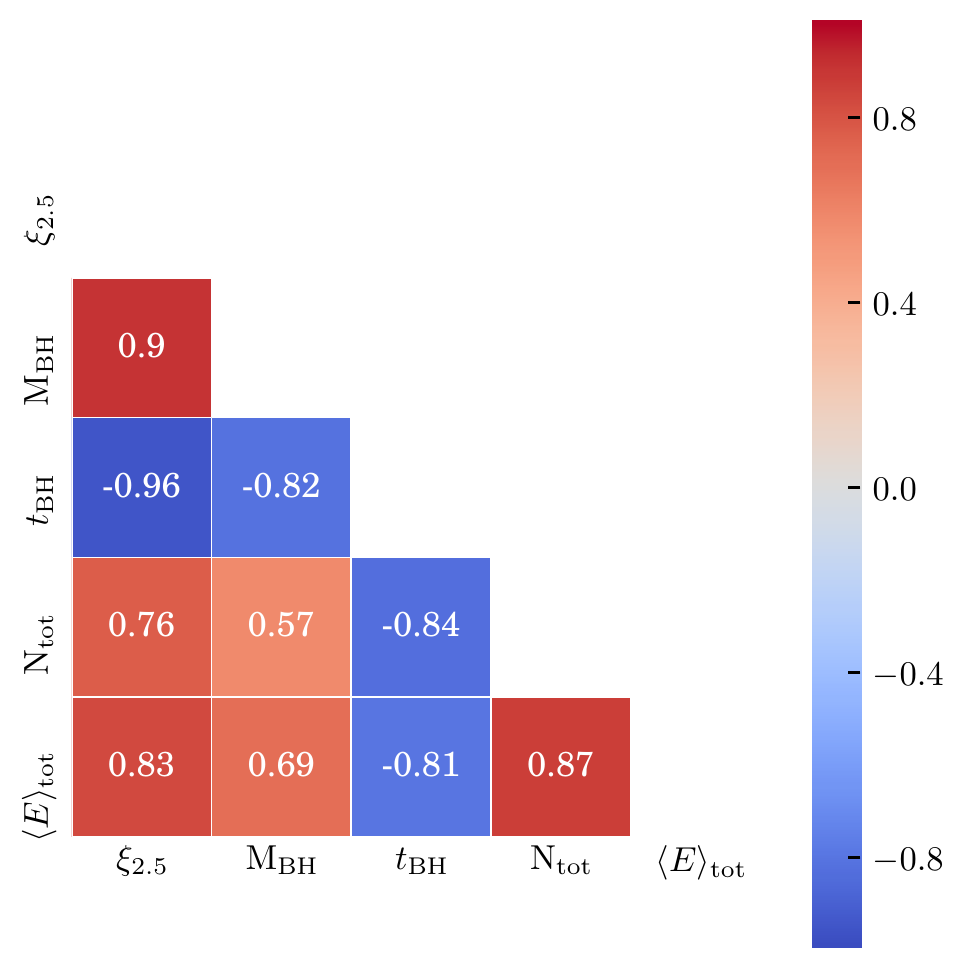}
\caption{Correlation matrix for observable neutrino quantities and underlying properties of the progenitor and collapse for simulations that do not explode and result in ``failed'' supernova events, leaving a remnant BH.  The quantities considered are:  core compactness ($\xi_{2.5}$), time to collapse to BH ($t_\mathrm{BH}$), time- and energy-integrated counts in all channels out to 1s post-bounce ($\mathrm{N}_\mathrm{tot}$), time-integrated average neutrino energy for all channels ($\langle E \rangle_\mathrm{tot}$) out to 1s post-bounce.  The lower-left half shows the correlation strength, ranging from -1 to 1, as indicated by the color and number in each of the squares. }
\label{fig:neut_bh}
\end{figure}

We perform the same analysis for simulations that fail to explode, as is shown in Figure~\ref{fig:neut_bh}.  This shows the correlation strength between the observable neutrino quantities, the BH mass, $\mathrm{M}_\mathrm{BH}$, the core compactness, $\xi_{2.5}$, and the time of collapse to BH, $t_\mathrm{BH}$.  We consider the time of collapse to BH an ``observable'' quantity for the sake of this analysis since it will be easily measurable from the duration of the neutrino emission, from the neutronization burst at bounce to the end of neutrino emission with collapse to a BH.  The BH mass was found using the approach of \cite{fernandez:2018}, which accounts for the unbinding of some of the stellar envelope due to neutrino losses before the collapse to a BH occurs. 

Surprisingly though, we find very different relationships between the progenitor properties and neutrino signal for failed supernovae than for successful explosions.  In the case of failed explosions, the neutrino mean energy is a better predictor of the core compactness than the neutrino counts.  The relationship between the core compactness and time to collapse to BH evident in Figure~\ref{fig:neut_bh} has been seen in previous works, first noted by  \cite{oconnor:2011} in studying failed CCSNe in 1D models.  The time to collapse to BH is strongly related to the free fall time scale of the inner $2.5 \,\mathrm{M}_{\odot}$ of the progenitor.

\begin{figure}[t]
\centering
\includegraphics[width = 0.5\textwidth]{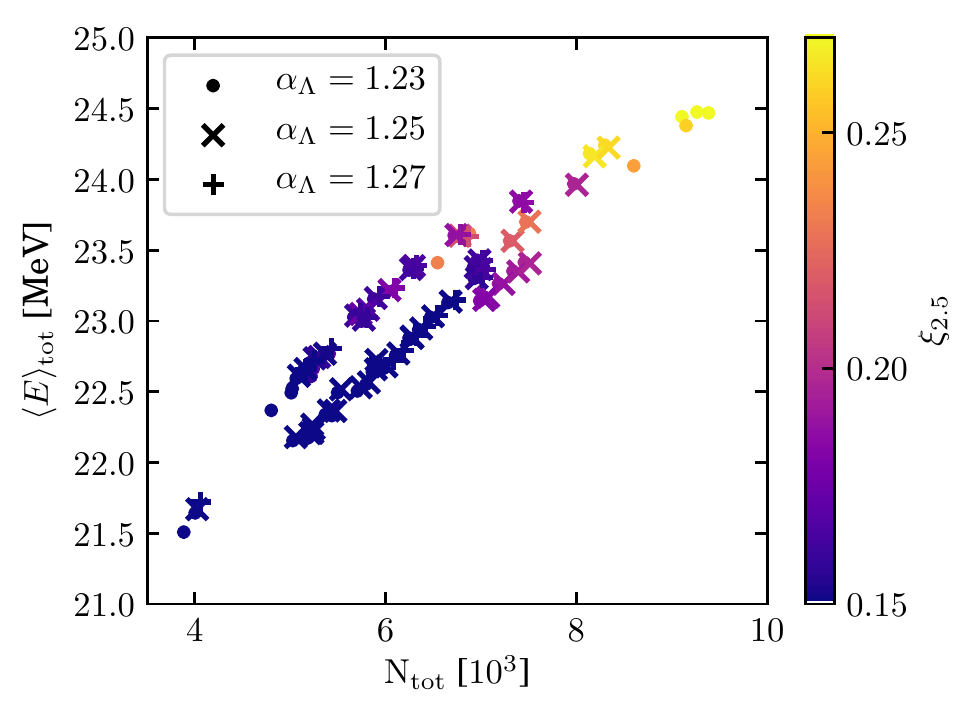}
\caption{The plane of  neutrino mean energy ($\langle E \rangle_\mathrm{tot}$) of the time-integrated observed neutrino spectrum versus time- and energy-integrated counts ($\mathrm{N_{tot}}$) in all channels out to 1s post-bounce, for failed events that collapse to a BH.   The color indicates the core compactness of the progenitor star ($\xi_{2.5}$). Symbols indicate the turbulence strength parameter $\alpha_{\Lambda}$ used in the simulation.   Detections are simulated for a 32 kton water Cherenkov detector for a CCSN event at $10 \,\mathrm{kpc}$. Given the lack of scatter with turbulence strength, this measurement appears to be insensitive to the assumptions made in our 1D supernova model. }
\label{fig:neut-bhcomp}
\end{figure}

Similar to our analysis for the successful explosions, observations of the neutrino count rates and average energies can be used to obtain the core compactness of stars that fail to explode.  Figure~\ref{fig:neut-bhcomp} shows the plane of neutrino average energies and counts for all of our simulations that collapse to a BH, with the core compactness indicated by the color.  As with the successful explosions, as shown in Figure~\ref{fig:neut-nsmass}, these quantities are tightly correlated and show little scatter between different values of the turbulence strength parameter $\alpha_{\Lambda}$.  Thus, the progenitor core compactness of a detected failed supernova event could be determined with measurements of the neutrinos out to $1\,\mathrm{s}$ post-bounce.  The coefficients of the linear fit of these data are shown in Table~\ref{tab:fail_neut_fits}, along with a fit for the BH mass.  These fits assume a functional form of 
\begin{equation}
y= \mathrm{C_{{t_{BH}}}} \mathrm{{t_{BH}}} +  \mathrm{C_\mathrm{N_{tot}}} \mathrm{N_{tot}} +  \mathrm{C_{\langle E \rangle_{tot}}} \mathrm{\langle E \rangle_{tot}} + \mathrm{D}~,
\end{equation}
where $ \mathrm{{t_{BH}}}$ is the time post-bounce to collapse to BH, $\langle E\rangle_\mathrm{tot}$ is the mean energy, and  $\mathrm{N_{tot}}$ is the number of neutrino counts.
We also include the coefficients for a fit that does not include the time to collapse to BH.  Although this fit is not quite as good as that including the time to collapse to BH,  knowing only the neutrino parameters out to $1\,\mathrm{s}$ with no knowledge of the collapse time will be sufficient for constraining the core compactness and BH mass.

\begin{table*}[t]
   \centering
   \topcaption{Linear fit coefficients for observable neutrino parameters for failed supernova events. Included are fit parameters if time to BH formation is unknown. Coefficients are for functional fit of the form $ y= \mathrm{C_{{t_{BH}}}} \mathrm{{t_{BH}}} +  \mathrm{C_\mathrm{N_{tot}}} \mathrm{N_{tot}} +  \mathrm{C_{\langle E \rangle_{tot}}} \mathrm{\langle E \rangle_{tot}} + \mathrm{D}$, with the time to collapse to BH, $t_\mathrm{BH}$, measured in seconds and the mean energy, $\langle E\rangle_\mathrm{tot}$, measured in MeV.  $R^{2}$ is the correlation coefficient between the data and the fit, and measures the quality of the fit.  Note that these fits assume a distance of $10\,\mathrm{kpc}$ and thus the number of counts will have to be scaled accordingly for CCSN events at other distances.  } 
   \begin{tabular}{@{} lccccc@{}} 
      \toprule
      Parameter & C$_\mathrm{t_{BH}}$ & C$_\mathrm{\langle E \rangle_{tot}}$ & C$_\mathrm{N_{tot}}$ & D  & $R^{2}$\\
      \midrule
     $\xi_{2.5}$ [unitless] &     $(-3.634 \pm 0.278) \times 10^{-2}$ & $(1.572 \pm 0.604) \times 10^{-2}$ & ---& $(-0.412 \pm 1.507)\times 10^{-1}$ & 0.927 \\
   & --- & $(9.025 \pm 0.336) \times 10^{-2}$ & --- & $-1.929 \pm 0.0775$ & 0.653 \\
      & $(-4.225 \pm 0.322) \times 10^{-2}$ & --- & $(0.000 \pm 3.233 ) \times 10^{-6}$ & $(3.507 \pm 0.338) \times 10^{-1}$ & 0.918 \\
   & --- & --- & $(4.409 \pm 0.275) \times 10^{-5}$ & $(-1.261 \pm 0.175) \times 10^{-1}$ & 0.841 \\
     $\mathrm{M}_\mathrm{BH}$ [$\mathrm{M}_{\odot}$] &  $-1.654\pm 0.291$ & $0.356 \pm 0.631$ & --- & $3.545 \pm 15.744$ & 0.676\\
      & --- & $3.268 \pm 0.465$ & --- & $-71.084 \pm 10.961$ & 0.478\\
     & $1.787 \pm 0.318$ & --- & $(0.000 \pm 3.191) \times 10^{-4}$ & $(1.241 \pm 0.333) \times 10^{1}$ & 0.674\\
       & --- & --- & $(1.257 \pm 0.244) \times 10^{-3}$ & $-3.120 \pm 1.773$ & 0.330 \\
       \bottomrule
   \end{tabular}
   \label{tab:fail_neut_fits}
\end{table*}

It is clear that just one second of neutrino detection from the next galactic supernova will provide a wealth of information, and with measurements from a just single detector.  Using a network of detectors, with different sensitivities and overall increased count rate, the correlations and constraints shown here should only improve.  It is important to note, however, that these results assume an event distance of $10\,\mathrm{kpc}$.  For events much further away with very low count rates, or with significant errors in the distance measurement, the statistical error in the number of counts could become overwhelming and prevent identification of the properties explored here.

Additionally, we do not consider  sensitivities to the nuclear equation of state.  It is well established that the neutrino emission and the time to collapse to BH will be sensitive to the equation of state properties \citep{oconnor:2011,yasin:2020,aschneider:2019}.  Many of these sensitivities, such as those to the symmetry energy and its slope, do not appear in the neutrino emission until many seconds post-bounce \citep{roberts:2012,aschneider:2019} and will not impact the results found here.  However, as shown by \cite{aschneider:2019}, the effective nucleon mass will impact on the neutrino emission in the first second.  We leave the exploration of the sensitivity of our parameter estimations to the equation of state to future work. 

\subsection{Gravitational waves\label{sec:gw}}

The analysis of the GW emission we perform here provides the peak frequency of the GW emission, but does not provide information about the strength of that emission.  Although 2- and 3-dimensional simulations provide more detailed and accurate information about the GW emission, it is still beneficial to consider 1D simulations since we can run more 1D simulations (600 in this paper) and can run them to later times than multidimensional simulations.

There has been much previous work that has considered the progenitor dependence of the GW emission \citep{murphy:2009,mueller:2013,yakunin:2015,morozova2018}.  These works considered the GW signal from 2D simulations of a few progenitor models and largely came to the same conclusions: the peak frequency is not strongly dependent on the progenitor, but the GW energy is much more strongly related to the progenitor.  As we will see, we find a similar result here, in that the peak GW frequency is more weakly correlated with the progenitor properties than  the neutrino emission, for example, and is not at all correlated with the ZAMS mass.  However, many of these studies did not consider the evolution of the peak GW frequency beyond $1\,\mathrm{s}$ post-bounce.  We find that there is little progenitor dependence at early times, but the peak GW frequencies begin to diverge at later times for different progenitors (see for example the bottom panel of Figure~\ref{fig:explvfail}).  

\begin{figure}[b]
\centering
\includegraphics[width = 0.5\textwidth]{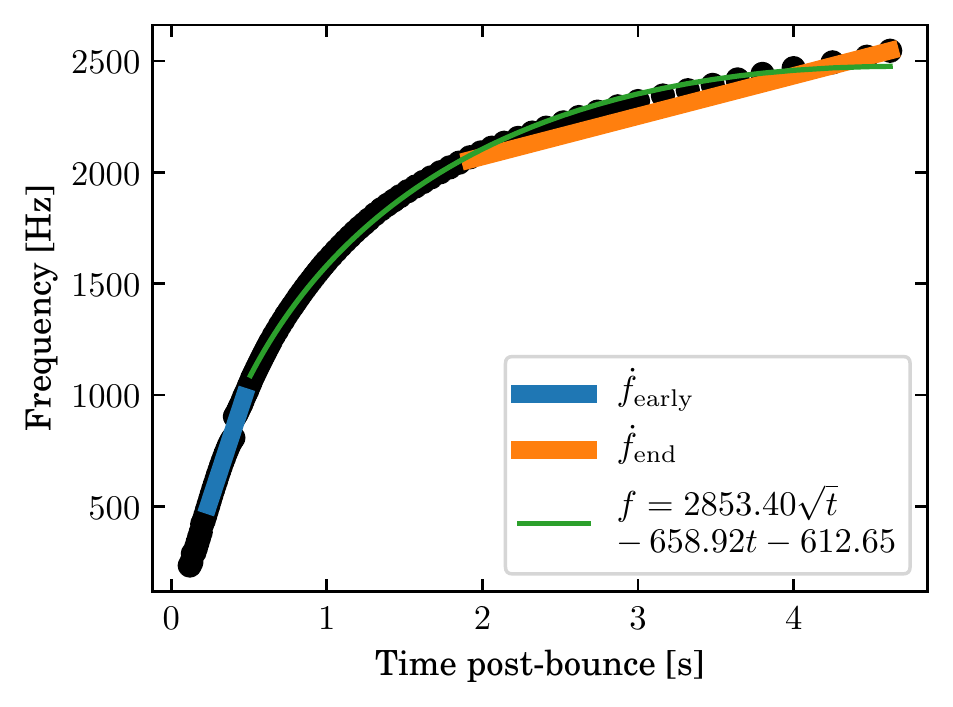}
\caption{Example of GW quantities considered in the correlation matrix for a  $15\,\mathrm{M}_{\odot}$ progenitor with $\alpha_{\Lambda} = 1.25$, which does not explode successfully.  The blue line is the early slope $\dot{f}_{\mathrm{early}}$ between $500$ and $1000 \,\mathrm{Hz}$ and the orange line shows the slope just prior to collapse $\dot{f}_\mathrm{end}$, between  collapse to BH and $500 \,\mathrm{Hz}$ below the final frequency.   These quantities will be used to compare between simulations and explore correlations with properties of the progenitor and explosion. Green line is the function fit of peak gravitational wave frequency after $0.5\,\mathrm{s}$ post-bounce.  }
\label{fig:GWexample}
\end{figure}

 In order to quantify the evolution of the dominant GW frequency and easily compare between a large number of simulations, we consider scalar quantities derived from the evolution of the peak frequency: the early slope of the frequency evolution $\dot{f}_{\mathrm{early}}$ from $500$ to $1000 \,\mathrm{Hz}$ and additionally for simulations that result in collapse to a BH we consider the slope of the frequency just prior to collapse, $\dot{f}_\mathrm{end}$, from collapse to BH to $500 \,\mathrm{Hz}$ below the frequency at the time of collapse.  We chose to use frequencies, rather than times post-bounce, as references for defining these slopes since they will be easier to determine from observational data.   For non-rotating CCSNe, it will be difficult to determine the time of bounce from the GW signal alone \citep{pajkos:2019}.  However, even with current detector resolution, it should be possible to distinguish the time between a $500 \,\mathrm{Hz}$ frequency change. Figure~\ref{fig:GWexample} shows an example of the dominant GW frequency evolution from a 1D simulation, for a $15\,\mathrm{M}_{\odot}$ progenitor with $\alpha_{\Lambda} = 1.25$, as well as the above parameters for this particular simulation.  For the same reason, we do not consider the time to collapse to a BH, $t_\mathrm{BH}$, to be an ``observable'' feature from the GW emission alone, since we will not have a reliable indicator of the time of bounce from GWs alone.  Additionally, because they are linear, these slopes are not strictly dependent on the specific windows listed here -- they could be computed using any available window depending on the available observed data.   However, it is worth noting that the end frequency may be difficult to detect as LIGO is not currently very sensitive to frequencies above $\sim 1000 \mathrm{Hz}$. 

 Although the parameters we consider seem rudimentary, we experimented with using various arbitrary functional fits to the frequency evolution with time, but found that as with the neutrino evolution, the fitting parameters were not well correlated with progenitor and explosion process, especially in comparison with the parameters we have adopted.   For example, \cite{morozova2018} used a quadratic fit of the form $f = A t^{2} + B t + C$ to characterize the peak GW frequency between $0.5\,\mathrm{s}$ and $1.5\,\mathrm{s}$ post-bounce.   However we have found that, although this functional fit well captures the shape of the peak frequency at early times, it fails to capture the behavior as the peak frequency levels out at late times ($t > 1.5\,\mathrm{s}$ post-bounce).  We propose instead a fit of the form $f = A \sqrt{t} + B t +C$, which better captures the behavior at later times. An example of such a fit is shown in Figure~\ref{fig:GWexample}. The coefficients of  both functional fits are not well correlated with properties of the progenitor or explosion. Nonetheless, such fits may be useful for studies of detectability, as priors in LIGO data searches, etc, and thus the fit coefficients can be found online,\footnote{Publicly available at \url{https://doi.org/10.5281/zenodo.3822556}} but we will not discuss them further in this work in our exploration of correlated quantities. 

\begin{figure}[t]
\centering
\includegraphics[width = 0.5\textwidth]{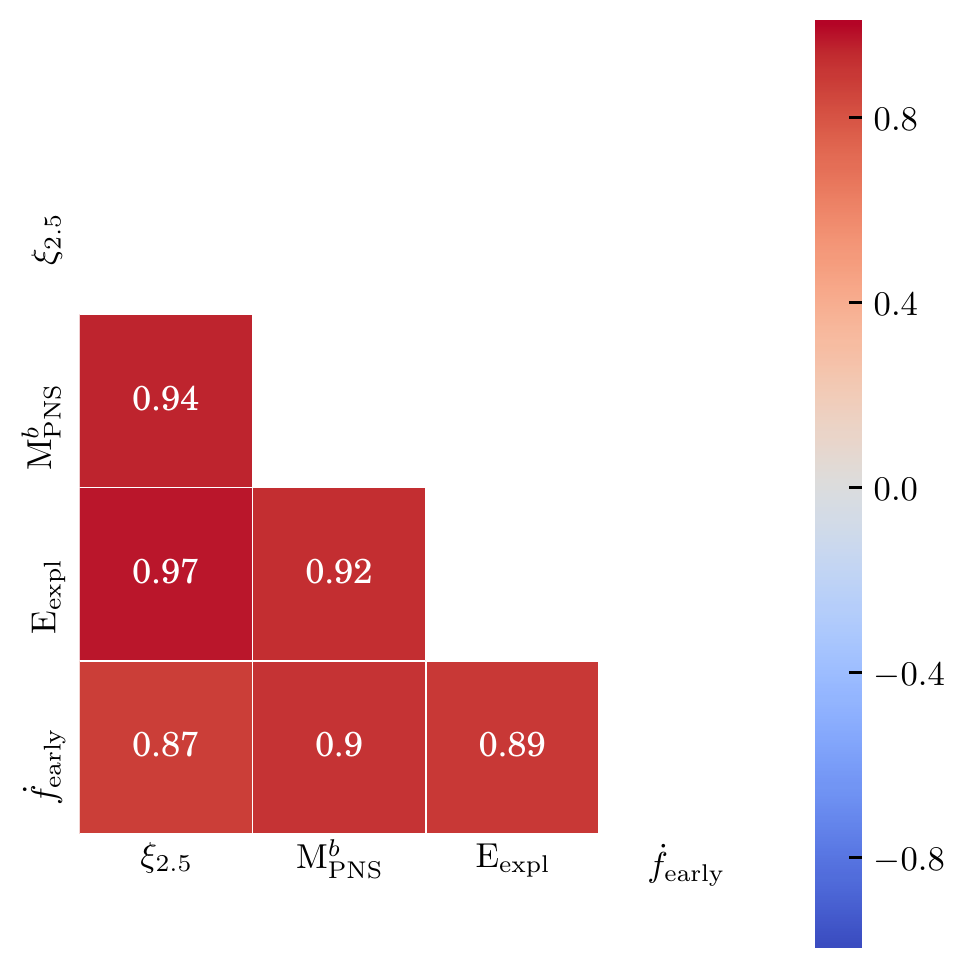}
\caption{Correlation matrix for observable GW quantities and underlying progenitor and explosion properties for simulation that successfully explode.  The quantities considered are:   core compactness ($\xi_{2.5}$), PNS mass ($\mathrm{M}^{b}_\mathrm{PNS}$) at $1\,\mathrm{s}$ post-bounce, explosion energy ($\mathrm{E_{expl}}$), early peak GW frequency slope ($\dot{f}_{\mathrm{early}}$) between $500$ and $1000 \,\mathrm{Hz}$.  The lower-left half shows the correlation strength, ranging from -1 to 1.  Red colors indicate a strong positive correlation and blue a negative correlations, while grey indicates no correlation.  The correlation between two values is also indicated by the number in each box. }
\label{fig:gw_ns}
\end{figure}

It is easiest to see how these GW parameters change with properties of the progenitor and explosion by looking at the correlation matrix.  Figure~\ref{fig:gw_ns} shows the correlation matrix between progenitor, explosion, and GW properties for simulations that explode successfully.  Again, there are many additional parameters that could be included, but all showed weak correlations with the GW emission and thus we do not include them here.  The GW eigenfrequencies are intimately related to the PNS mass and the core compactness $\xi_{2.5}$, since they are sensitive to the structure of the core and its evolution.

\begin{figure}[b]
\centering
\includegraphics[width = 0.5\textwidth]{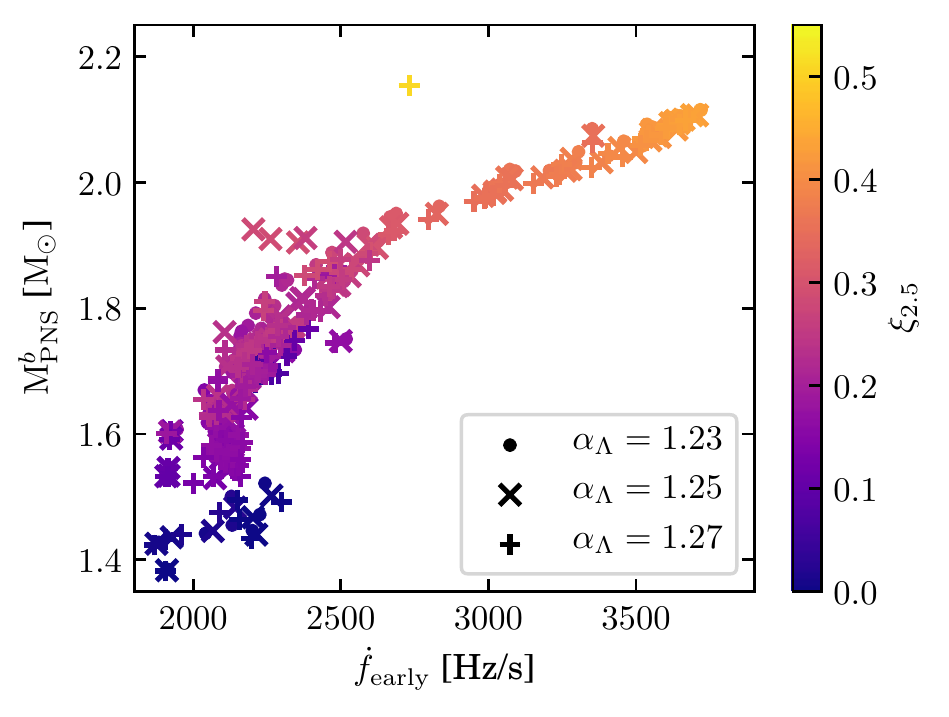}
\caption{The PNS baryonic mass ($\mathrm{M}^{b}_\mathrm{PNS}$)   at $1\,\mathrm{s}$ post-bounce versus  slope of the peak GW frequency ($\dot{f}_{\mathrm{early}}$) between $500$ and $1000 \,\mathrm{Hz}$, for simulations that explode successfully.  The core compactness ($\xi_{2.5}$) is indicated by the color.  The turbulence strength parameter $\alpha_{\Lambda}$ is indicated by the symbol. }
\label{fig:gw-nsmass}
\end{figure}

The PNS mass, $\mathrm{M}^{b}_\mathrm{PNS}$, at $1\,\mathrm{s}$ post-bounce is strongly correlated with the early time evolution of the dominant GW frequency $\dot{f}_{\mathrm{early}}$.  Thus, similar to what we have done with the neutrino emission, the PNS mass can be constrained by the GW emission.  Figure~\ref{fig:gw-nsmass} shows the PNS mass versus the early time evolution of the peak GW frequency $(\dot{f}_{\mathrm{early}})$.  There is a less clear trend here than for the neutrino counts, which is to be expected given the weaker correlations between the GW emission and PNS mass than for the neutrino counts.  Nonetheless, detection of the early GW signal will again provide an independent measure of the PNS mass before fallback occurs.  Table~\ref{tab:expl-gw-fits} contains the linear fit coefficients for these parameters and the core compactness, PNS mass, and explosion energy.  These fits use a functional form
\begin{equation}
y= \mathrm{C_{\dot{f}_\mathrm{early}}} \dot{f}_\mathrm{early} + \mathrm{D}~,
\end{equation}
where $\dot{f}_\mathrm{early}$ is the early time evolution of the peak GW frequency.
Note that there is clearly a nonlinear relationship between these variables, and a nonlinear fit may provide a stronger correlation and fit estimation.

\begin{table*}[t]
   \centering
   \topcaption{Linear fit coefficients for observable GW parameters for successful supernova events for the core compactness ($\xi_{2.5}$), baryonic PNS mass at $1\,\mathrm{s}$ post-bounce ($\mathrm{M}^{b}_\mathrm{PNS}$), and explosion energy ($\mathrm{E_{expl}}$).  Coefficients are for functional fit of the form $ y= \mathrm{C_{\dot{f}_\mathrm{early}}} \dot{f}_\mathrm{early} + \mathrm{D}$, with $\dot{f}_\mathrm{early}$ measured in $\mathrm{Hz/s}$. $R^{2}$ is the correlation coefficient between the data and the fit, and measures the quality of the fit.  } 
   \begin{tabular}{@{} lcc c  @{}} 
      \toprule
      Parameter &   C$_{\dot{f}_\mathrm{early}}$ &  D & $R^{2}$\\
      \midrule
     $\xi_{2.5}$ [unitless] &   $(1.790 \pm 0.047) \times 10^{-4}$ & $(-2.110 \pm 0.118) \times 10^{-1}$  & 0.763  \\
     $\mathrm{M}^{b}_\mathrm{PNS}$ [$\mathrm{M}_{\odot}$] &   $(3.131 \pm 0.070) \times 10^{-4}$ & $(9.945 \pm 0.175) \times 10^{-1}$ & 0.818 \\
     $\mathrm{E_{expl}}$ [$10^{51}\,\mathrm{erg}$] & $(6.848 \pm 0.169) \times 10^{-4}$ & $(-7.880 \pm 0.427) \times 10^{-1}$ & 0.784 \\
       \bottomrule
   \end{tabular}
   \label{tab:expl-gw-fits}
\end{table*}

\begin{figure}[b]
\centering
\includegraphics[width = 0.5 \textwidth]{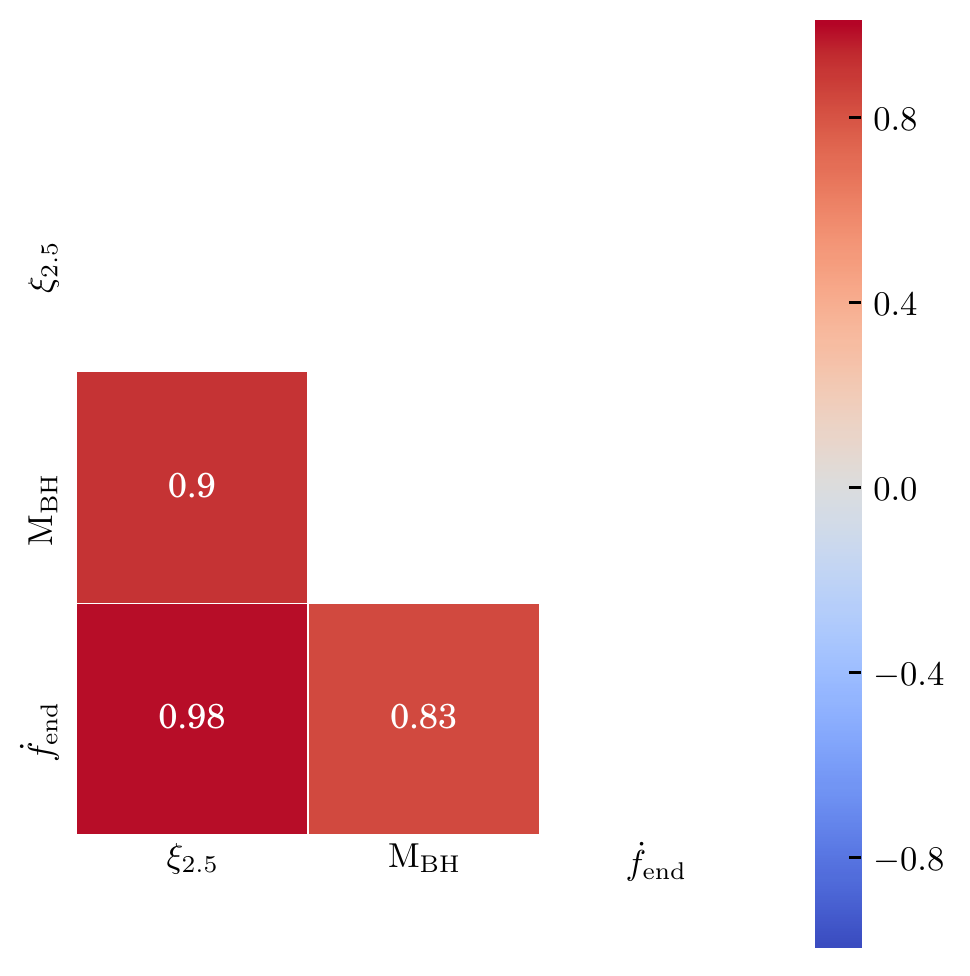}
\caption{Correlation matrix for observable GW quantities and underlying progenitor for simulations that do not explode and result in ``failed'' supernova events.   The quantities considered are:  core compactness ($\xi_{2.5}$), BH mass ($\mathrm{M_{BH}}$), and late GW frequency slope ($\dot{f}_\mathrm{end}$) just prior to collapse to BH. The lower-left half shows the correlation strength, ranging from -1 to 1. }
\label{fig:gw_bh}
\end{figure}

Similarly, we can consider the relationship between progenitor and GW properties for failed explosions, as shown in Figure~\ref{fig:gw_bh}.  We consider the core compactness $\xi_{2.5}$ and BH mass $\mathrm{M}_\mathrm{BH}$.  Again, we do not consider the time to collapse to BH an ``observable'' quantity, for the sake of this analysis, since it will not be easily measurable from the duration of the GW emission. A tight connection has also been noted between the core compactness and BH formation time was pointed out by \cite{oconnor:2011}, who connected the BH formation time with the free-fall timescale of the 2.5 $\mathrm{M}_{\odot}$ mass shell.  It may be possible to discern the BH formation time indirectly from the compactness.  

\begin{figure}[b]
\centering
\includegraphics[width = 0.5\textwidth]{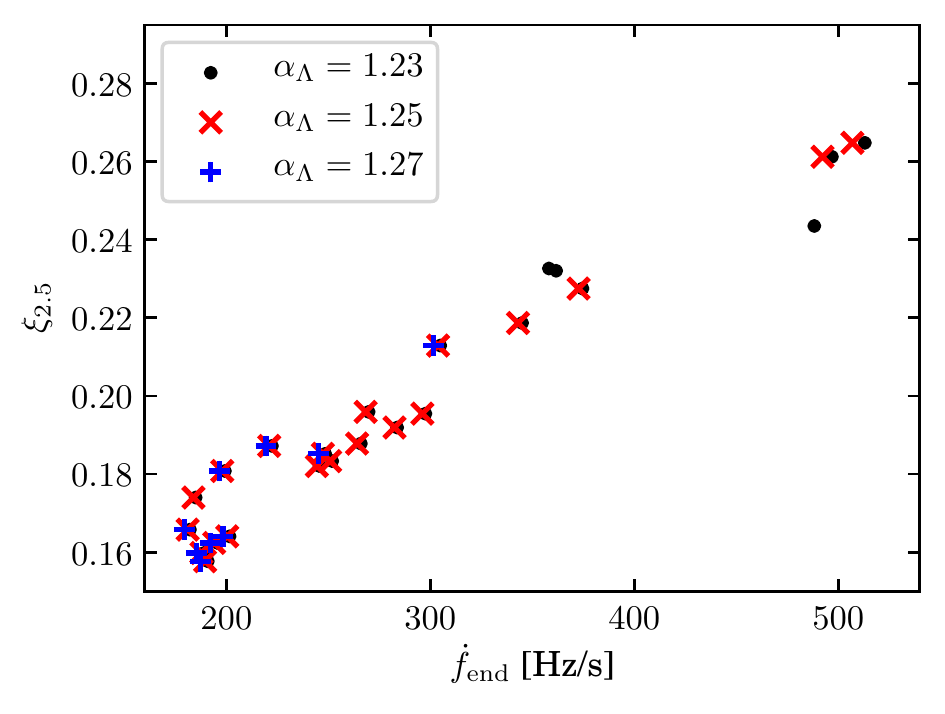}
\caption{Core compactness, $\xi_{2.5}$, versus the slope of the late time dominant GW frequency, $\dot{f}_\mathrm{end}$,  for simulations that fail to explode.  The turbulence strength parameter $\alpha_{\Lambda}$ is indicated by the color and symbol.}
\label{fig:gw-bhcomp}
\end{figure}

Surprisingly, or perhaps unsurprisingly after our discussion of the neutrino emission, there is a stronger relationship between the core compactness $\xi_{2.5}$ and the GW emission here than for successful explosions. The late time evolution of the dominant frequency $\dot{f}_{\mathrm{end}}$ just prior to collapse to BH correlates strongly with the core compactness. Figure~\ref{fig:gw-bhcomp} shows the core compactness versus  the evolution of the dominant GW frequency just prior to collapse to BH.  There is a clear, distinct trend in core compactness and GW emission.  Table~\ref{tab:fail_gw_fits} provides the fits between the observable GW emission and the core compactness and BH mass for core collapse events that fail to explode, assuming a function of the form
\begin{equation}
 y=  \mathrm{C_{\dot{f}_\mathrm{end}}}\dot{f}_\mathrm{end} + \mathrm{D}~,
 \end{equation}
 where $\dot{f}_\mathrm{end}$ is the late time evolution of the dominant GW frequency.

\begin{table*}[t]
   \centering
   \topcaption{Linear fit coefficients for observable GW parameters from failed supernova events for the core compactness ($\xi_{2.5}$) and BH mass ($\mathrm{M_{BH}}$).  Coefficients are for functional fit of the form $ y=  \mathrm{C_{\dot{f}_\mathrm{end}}}\dot{f}_\mathrm{end} + \mathrm{D}$, with $\dot{f}_\mathrm{end}$ measured in $\mathrm{Hz/s}$.  $R^{2}$ is the correlation coefficient between the data and the fit, and measures the quality of the fit.} 
   \begin{tabular}{@{} lc c c  @{}} 
      \toprule
      Parameter & C$_{\dot{f}_\mathrm{end}}$ & D & $R^{2}$ \\
      \midrule
     $\xi_{2.5}$ [unitless] &  $(2.873 \pm 0.083) \times 10^{-4}$ & $(1.143 \pm 0.027) \times 10^{-1}$ & 0.957 \\
     $\mathrm{M}_\mathrm{BH}$ [$\mathrm{M}_{\odot}$] & $(1.208\pm0.109) \times 10^{-2}$ & $2.428 \pm 0.347$ & 0.694\\
       \bottomrule
   \end{tabular}
   \label{tab:fail_gw_fits}
\end{table*}

In summary, the GW emission provides an additional window into the central engine of CCSNe.  With detections of \textit{only} the dominant GW frequency, we will be able to place tight constraints on the progenitor star and remnant compact object.   We only consider the GW emission from eigenmodes of the PNS, and not from the convective material near the shock, which may encode additional information about the progenitor \citep{torres-forne}.

\subsection{Combined analysis\label{sec:corr}}

We have shown that detection of \textit{either} neutrinos or GWs from a nearby CCSN will provide constraints on the stellar structure, mass of the PNS and BH prior to fallback, and explosion energy.  However, it is likely that we will have detections of {\it both} GWs and neutrinos from such an event.  For successful and failed explosions, neutrino and GW quantities are both strongly correlated with the same properties of the progenitor and explosion.  Therefore, combining these detections may provide an even more robust way to constrain the progenitor and explosion properties of interest.

\begin{table*}[t]
   \centering
   \topcaption{Linear fit coefficients to observable neutrino and GW parameters for successfully exploding CCSN events for the core compactness ($\xi_{2.5}$), PNS mass at 1s post-bounce ($\mathrm{M}^{b}_\mathrm{PNS}$), and explosion energy ($\mathrm{E}_\mathrm{expl}$).  Coefficients are for functional fit of the form $ y= \mathrm{C_{N_{tot}}} \mathrm{N_{tot}} +  \mathrm{C_{\langle E \rangle_{tot}}} \langle E \rangle_\mathrm{{tot}} +  \mathrm{C_{\dot{f}_{early}}} \dot{f}_\mathrm{early} + \mathrm{D}$, with the mean energy, $\langle E \rangle_\mathrm{tot}$, measure in MeV and $\dot{f}_\mathrm{early}$ measured in $\mathrm{Hz/s}$.  $R^{2}$ is the correlation coefficient between the data and the fit, and measures the quality of the fit. Note that these fits assume a distance of $10\,\mathrm{kpc}$ and thus the number of counts will have to be scaled accordingly for CCSN events at other distances. } 
   \begin{tabular}{@{} lcc c c c c @{}} 
      \toprule
      Parameter & C$_\mathrm{N_{tot}}$ & C$_{\langle E \rangle_\mathrm{tot}}$ &  C$_{\dot{f}_\mathrm{early}}$ &  D  & $R^{2}$ \\
      \midrule
     $\xi_{2.5}$ [unitless] &   $(3.771 \pm 0.409) \times 10^{-5}$ & --- & $(1.130 \pm 0.083) \times 10^{-4}$ & $(-2.034 \pm 0.109) \times 10^{-1}$ & 0.801\\
     					 &  --- & $(1.926 \pm 0.487) \times 10^{-2}$& $(1.637 \pm 0.060 ) \times 10^{-4}$ & $(-5.690 \pm 0.912) \times 10^{-1}$ & 0.771   \\
     $\mathrm{M}^{b}_\mathrm{PNS}$ [$\mathrm{M}_{\odot}$] &  $(1.152\pm 0.038) \times 10^{-4}$ &---&  $(1.126 \pm 0.076) \times 10^{-4}$  & $1.018 \pm 0.010$ & 0.941   \\
      											 & --- & $(9.634 \pm 0.058) \times 10^{-2}$& $(2.388 \pm0.070) \times 10^{-4}$ & $(-7.959 \pm 1.079) \times 10^{-1}$ &  0.888 \\
     $\mathrm{E}_\mathrm{expl}$ [10$^{51}$ erg] &  $(1.291 \pm 0.149) \times 10^{-4}$ & --- &  $(4.600 \pm 0.303) \times 10^{-4}$ & $(-7.620 \pm 0.397) \times 10^{-1}$  & 0.815\\
     &  --- & $(7.161 \pm 1.754 ) \times 10^{-2}$   & $(6.295 \pm 0.215) \times 10^{-4}$& $-2.119 \pm 0.329$ &0.792  \\
       \bottomrule
   \end{tabular}
   \label{tab:expl_all_fits}
\end{table*}

For successful explosions, both the neutrino counts and GW evolution are correlated with the core compactness, $\xi_{2.5}$, PNS mass at 1s post-bounce, $\mathrm{M}^{b}_\mathrm{PNS}$, and explosion energy $\mathrm{E}_\mathrm{expl}$, as shown in Figures~\ref{fig:neut_ns} and \ref{fig:gw_ns}.  Table~\ref{tab:expl_all_fits} contains the linear fit coefficients for a combined measurement using both neutrinos and GWs, using a function of the form
\begin{equation}
y= \mathrm{C_{N_{tot}}} \mathrm{N_{tot}} +  \mathrm{C_{\langle E \rangle_{tot}}} \langle E \rangle_\mathrm{{tot}} +  \mathrm{C_{\dot{f}_{early}}} \dot{f}_\mathrm{early} + \mathrm{D}~,
\end{equation}
where $\mathrm{N_{tot}}$ is the number of neutrino counts, $\langle E \rangle_\mathrm{{tot}}$ is the neutrino mean energy, and $\dot{f}_\mathrm{early}$ is the early time evolution of the dominant GW frequency.  By simultaneously using neutrino and GW information, there is an improvement in the quality of the fit, as measured by the $R^{2}$, for both the core compactness and PNS mass. For example, the $R^{2}$ of the fit for the core compactness was $0.841$ using just the neutrino counts and $0.763$ using the GW slope, but is increased to $0.801$ when using both neutrino counts and the GW information. The addition of GW information to the neutrino information may prove to be an invaluable addition to constraining the core compactness, PNS mass, and explosion energy for CCSN events at large distances with low count rates or for events with large uncertainties on the distance.

\begin{table*}[t]
   \centering
   \topcaption{Linear fit coefficients to observable neutrino and GW parameters for failed supernova events for the core compactness ($\xi_{2.5}$) and BH mass ($\mathrm{M_{BH}}$).  Included are fit parameters if time to BH formation is unknown.  Coefficients are for functional fit of the form $ y= \mathrm{C_{t_{BH}}} \mathrm{t_{BH}} + \mathrm{C_{\dot{f}_\mathrm{end}}}\dot{f}_\mathrm{end} + \mathrm{D}$, with the time to collapse to BH, $t_\mathrm{BH}$, measured in seconds, and $\dot{f}_\mathrm{end}$ measured in Hz/s.  $R^{2}$ is the correlation coefficient between the data and the fit, and measures the quality of the fit. Note that these fits assume a distance of $10\,\mathrm{kpc}$ and thus the number of counts will have to be scaled accordingly for CCSN events at other distances. } 
   \begin{tabular}{@{} lcc ccc @{}} 
      \toprule
      Parameter & C$_{t_\mathrm{BH}}$ &   C$_{\dot{f}_\mathrm{end}}$ & D  & $R^{2}$ \\
      \midrule
     $\xi_{2.5}$ [unitless] & $(-1.074 \pm0.429) \times 10^{-2}$ & $(2.186\pm 0.286) \times 10^{-4}$ & $(1.732 \pm0.236) \times 10^{-1}$ & 0.961 \\
       & --- &  $(2.873\pm0.083)\times 10^{-4}$  & $(1.114 \pm 0.026) \times 10^{-1}$ & 0.957 \\
       $\mathrm{M}_\mathrm{BH}$ [$\mathrm{M}_{\odot}$]  & $(-5.837 \pm5.894) \times 10^{-1}$ & $(8.348 \pm 3.927) \times 10^{-3}$  &$5.626 \pm 3.248$ & 0.700 \\
     &--- &    $(1.208 \pm 0.109)\times 10^{-2}$   & $2.428 \pm 0.347$ & 0.694 \\
       \bottomrule
   \end{tabular}
   \label{tab:fail_all_fits}
\end{table*}

For failed supernova events, both the neutrino and GW observables are correlated with the core compactness, $\xi_{2.5}$, and remnant BH mass, $\mathrm{M}_\mathrm{BH}$.  For both of these parameters, the quality of the fit is \textit{not} noticeably improved by combining observables from both messengers. Table~\ref{tab:fail_all_fits} contains the fit coefficients for such a combined analysis, using a function of the form
\begin{equation}
y= \mathrm{C_{t_{BH}}} \mathrm{t_{BH}} + \mathrm{C_{\dot{f}_\mathrm{end}}}\dot{f}_\mathrm{end} + \mathrm{D}~,
\end{equation}
where $ \mathrm{t_{BH}}$ is the time to BH formation and $\dot{f}_\mathrm{end}$ is the late time evolution of the dominant gravitational wave frequency.  Including the neutrino mean energy $\langle E \rangle_\mathrm{tot}$ or total counts $N_\mathrm{tot}$ along with the gravitational wave information was not advantageous and over constrained the fit, leading to the coefficients on the neutrino data to be zero in the fit.  These fits are not included here.  There is a slight improvement in the quality of the fit if the time to BH formation, $t_\mathrm{BH}$, as could be measured using neutrinos, is included alongside the slope of the gravitational wave frequency $\dot{f}_\mathrm{end}$.  For the core compactness, the $R^{2}$ value increases to 0.961 from 0.957 if the time to black hole formation is included in the fit.  For failed CCSN events, it will perhaps be more significant to determine the core compactness and BH mass from the neutrino and GW signals separately, since they both independently provide tight constraints on these parameters, and ensure that they are in agreement.  Disagreement in core compactness and/or BH mass as determined from the GW and neutrino signals separately would be indicative that we do not fully understand the neutrino and GW emission from failed CCSNe.

\section{Discussion and Conclusions\label{sec:conclusions}}

Using 600 1D simulations of CCSNe from $9-120\,\mathrm{M}_{\odot}$, carried out with the STIR model in the FLASH supernova code, we analyze the observable neutrino and GW signals to study the correlations with properties of the progenitor star, explosion mechanism, and remnant compact object.  To model the neutrinos, we simulate the detected neutrino signal in a $32\,\mathrm{kton}$ Super Kamiokande-like water Cherenkov detector using the \snowglobes~code.  To model the GWs, we conduct an astroseismology analysis of the PNS structure to determine the dominant GW frequency.  

Given the significant computational expense of 3D simulations, it is not currently feasible to use 3D simulations to make predictions of the multi-messenger signals of the entire range of possible CCSN events.  1D simulations remain vital for performing parameter and population studies such as this.  Although these simulations are done in 1D, they have been shown to reproduce the features of 3D simulations.  The STIR model uses a Reynold averaged decomposition of the Euler equations to model the effects of turbulence in 1D simulations and reproduces the behavior of 3D simulations \citep{couch:2020}.  Additionally, the GW eigenfrequencies found here match the peak GW frequency seen in 3D simulations, as shown in Figure~\ref{fig:GWcomp}.  

To address the questions posed in the introduction: first, will we be able to distinguish between a ``successful'' versus ``failed'' CCSN event without direct detection of EM emission?  We find that, with measurement of the first few seconds of neutrino and GW emission from a nearby CCSN, we may be able to determine whether or not it will explode successfully before observing EM emission. The time of collapse to BH in failed CCSNe will be marked by a sharp end in both the neutrino and GW emission.  More significantly, it may be possible to determine the time that the explosion sets in from a drop in both the neutrino counts and mean energy, but with continued emission at a lower rate and mean energy, as shown in Figure~\ref{fig:explvfail}.  The drop in the neutrino count rate and mean energy is tied to the decrease in the mass accretion onto the PNS after explosion sets in, an effect which is significantly enhanced in 1D simulations as compared to multidimensional.  The same effect is visible in current multidimensional simulations, although less pronounced. 

Secondly, what can we tell about the progenitor star, explosion mechanism, and remnant compact object from detection of neutrinos, GWs, or both? Additional quantitative information about the progenitor and explosion can also be determined from just a few seconds of detection as well, such as a determination of the PNS and BH mass and core compactness.  The neutrino and GW emission are strongly correlated with these parameters.  Determining these quantities from the neutrino and GW emission will aid in placing constraints on the progenitor star, independent of pre-explosion imaging, the evolution of the compact object prior to fallback accretion, and the explosion energy, independent of the light curve.  If we can determine the core compactness of the next galactic CCSN with the neutrino and GW emission, we can then compare that quantity with the results of stellar evolution calculations and the stellar mass found from pre-explosion imaging to either validate or constrain current stellar evolution models. We provide fits to our simulations that connect the observable neutrino and GW parameters with the core compactness, explosion energy, PNS mass, and remnant BH mass.  Additionally, we are making the neutrino and GW emission from our simulations publicly available, in hopes that they can be used in future studies of detectability and optimization of detector signal processing.

Although previous works have claimed that there is little progenitor dependence in the GW signal \citep{murphy:2009,mueller:2013,yakunin:2015,morozova2018}, we have found that the evolution of in the GW signal is strongly correlated with the core compactness, remnant compact object mass, and explosion energy; for example, the core compactness and BH mass are tightly correlated with the dominant GW frequency for massive stars that do not explode, as shown in Figure~\ref{fig:gw_bh}.  Although  there are significant differences between progenitors in the overall magnitude of the peak GW frequency at a given time post-bounce (for example, see the bottom panel of Figure~\ref{fig:explvfail}), we have limited ourselves to considering the slope of the GW signal.  These differences in the overall magnitude should be well described by the differences in the early slope of the peak GW frequency, which we have explored here.  More work must be done to determine how well this result holds up in multidimensional simulations, as well as after considering proper detector response.   Multidimensional effects, such as accretion, rotation, and SASI, may impact the late time GW emission in ways that cannot be captured in spherically symmetric simulations.  Additionally, there will be significant information embedded in the strength of the GW signal, as well, which cannot be explored using 1D.  

The correlations and fits found here assume an event distance of $10\,\mathrm{kpc}$.  Thus, for the fits provided, the number of neutrino counts must be scaled appropriately for events at different distances.  We find that, at $10\,\mathrm{kpc}$, the errors on the number of counts will be small and allow for a reliable determination of parameters  such as the PNS mass, as shown in Figure~\ref{fig:neut-nsmass}.  However, for events at larger distances where the count rate is low, or if there is a large error associated with the distance determination, the number of neutrino counts (scaled to 10kpc for use in our fits) may have a larger error and make it preferable to use the mean energy or GW emission in the determination of the core compactness and other parameters.  We consider only the neutrino detections in a single neutrino detector -- Super Kamiokande.  The next galactic supernova will be detected in a multitude of detectors, all providing additional insight and information into the CCSN environment.  The different detector sensitivities and overall increased count rates will only increase the significance of the correlations and relationships found here and may provide new correlations and insights.

 We find that, for successful CCSN events, combining the neutrino and GW information will improve the determination of parameters such as the core compactness.  However, for failed CCSN events, combining the neutrino and GW information did not improve the quality of fit in the determining the core compactness or BH mass.  Both signals will independently be able to constrain these quantities quite well, though, and combining them in this case may be artificially over constraining the fit.   Even if both neutrinos and GWs are not detected, we see that measurement of just one ``messenger'' will still provide constraints on the stellar structure and remnant compact object. It is promising that, for the next nearby CCSN event, there are multiple avenues for using either the GW or neutrino emission, or both, to constrain the stellar structure and remnant compact object and doing so does not depend on getting a single, specific detection with high precision.

Similarly, we only consider the evolution of the dominant GW frequency.  There will be detections of GW emission in other frequency bands, which also contain information about the structure of the PNS and the convection and turbulence occurring above it that excite those modes.  Further work must be done to determine what other frequency bands may be detectable with current and near-future detector sensitivities and what information they encode about the CCSN environment. Additionally, we take care with the neutrino emission to explore only detectable features by using \snowglobes~to simulate a realistic detected signal, but no such code is publicly available to simulate the detector response and reconstruction for GW signals.  Hopefully, by focusing on the peak GW frequency and its slope over broad spans of time, we have chosen detectable features.

The framework for analysis of CCSN simulations presented here has many possible applications and avenues for expansion.  There is already a significant body of work that explores the relationship between the supernova light curve and explosion energy, nucleosynthesis, etc.  The approach taken in this work can be applied to studies of the light curve as well and may aid in uncovering connections between EM detections, other multi-messenger signals, and progenitor properties \citep{barker:2019}.  As we see here, combining information from different sources can prove to be more insightful than a single source alone.  Including EM emission in our correlation analysis may uncover connections between the core of the event, as detected in neutrinos and GWs, and the shock propagation and explosion, as detected in EM emission.  Additionally, there has been no systematic study of EM emission from failed supernovae \citep{lovegrove:2013,quataert:2019}.  Correlations between EM emission and other signals may prove to be just as significant for failed events as for successful explosions.

We limit ourselves to linear correlations between the properties in this work.  There are far more complex and interesting relationships between the progenitor, CCSN environment, and detectable quantities than can be fully captured by linear relationships.  It is, if anything, quite shocking how strong the linear correlations are.  Future work will loosen this constraint by considering non-linear correlations and more complex functional fits between observable quantities and progenitor and explosion properties.

Finally, there is significant and impactful physics that has not been considered here.  Uncertainties due to binarity, rotation, and equation of state will add uncertainty to the determination of progenitor and explosion properties from multi-messenger signals.  However, these properties of rotation, equation of state, etc., may also have their own strong correlations with information encoded in the GW and neutrino signals.  For example, \cite{schneider:2019} found a relationship between the rotation and GW and neutrino emission that may be encoded in correlations such as those explored here.  \cite{aschneider:2019}  found correlations between equation of state parameters and neutrino emission in a $20\,\mathrm{M}_{\odot}$ progenitor and \cite{richers:2017}  explored correlations between equation of state parameters and the GW signal at bounce for rotating CCSNe.  Thus, multi-messenger detections may be able to provide constraints on additional stellar properties not considered in this work, and perhaps could be used to place constraints on other aspects of the problem such as the nuclear equation of state.  We plan to explore the sensitivity of the correlations found in this work to the nuclear equation of state, and any possible correlations between multi-messenger signals and equation of state parameters, in future work \citep{barker2:2019}. 

Perhaps the most significant omission is that of neutrino flavor mixing. We fundamentally do not know what flavor mixing processes may occur in CCSNe, and under what conditions.  Flavor mixing processes such as collective oscillations \citep{duan2010} and fast oscillations \citep{sawyer:2005,sawyer:2009,martin:2020,morinaga:2020,johns:2020}  have only been studied using stability analysis or simplified conditions.  The occurrence of any of these processes in the CCSN environment would drastically change how the observed neutrino signal is related to the GW emission, progenitor properties, and explosion dynamics.   It is worth noting, however, that the information contained in the neutrino signal may not be ``lost'' if significant flavor mixing processes occur -- merely, in some sense, scrambled.

There is a wealth of information that will be embedded in the neutrino and GW emission of the next nearby CCSN event and, with modern neutrino and GW detectors, we are well prepared to make such a detection.  However, interpreting these observations will require understanding the emission from all possible CCSN events, spanning the range of progenitor mass, metallicity, rotation, etc., and understanding the impact of nuclear equation of state and other fundamental physics uncertainties.  Given the computational cost of 3D simulations, we must continue to rely on 1D simulations to understand the span of possible CCSNe for the foreseeable future.  The study we conduct here is a first step in the process of predicting multi-messenger signals from all possible CCSNe and developing a framework to interpret the large number of simulations required to understand the landscape of CCSN events.

\acknowledgments{The authors would like to thank Kate Scholberg, Gail Mclaughlin, and Jim Kneller for insightful conversations about this work and Jost Migenda for helping with the formatting of the publicly available neutrino fluxes.
MLW is supported by an NSF Astronomy and Astrophysics Postdoctoral Fellowship under award AST-1801844 and by the U.S. Department of Energy, Office of Science, Office of Nuclear Physics, under Award Number DE-SC0015904.
SMC is supported by the U.S. Department of Energy, Office of Science, Office of Nuclear Physics, under Award Numbers DE-SC0015904 and DE-SC0017955.
EO is supported by the Swedish Research Council under Award Number 2018-04575.
This work was supported in part by Michigan State University through computational resources provided by the Institute for Cyber-Enabled Research.}

\software{FLASH \citep{fryxell:2000,dubey:2009}, \snowglobes \citep{scholberg:2012}, astroseismology analysis code \citep{morozova2018}, Matplotlib \citep{hunter:2007}, Numpy \citep{vanderwalt:2011}, SciPy \citep{scipy}, NuLib \citep{oconnor:2015}}
\newline

\bibliography{multimessenger}

\appendix

\section{Parameters considered in correlations\label{sec:parameters}}

\begin{table}[h]
\centering
\topcaption{Parameters considered in correlations.  Quantities not included in the main text were found to have weak or no correlation with multi-messenger signals or were describing similar physics to parameters that were used. }
 \begin{tabular}{@{} l l  @{}} 
      \toprule
        Quantity & Parameters\\
      \midrule

	Stellar structure & ZAMS mass \\
				& mass at time of core collapse \\
				& stellar radius at time of core collapse \\ 
				& core compactness ($\xi_{2.5}$) \\
				& iron core mass \\
				& Ertl parameters $\mu$ and $\mathrm{M}_{4}$ \citep{ertl:2016}\\
	Explosion dynamics & Explosion energy \\
				& Mass accretion rate \\ 
				& Heating rate in the gain region \\
				& Time of the onset of explosion \\
				& PNS radius \\
	Neutrino signal & Maximum counts in on 5ms time bin\\
				& Timescale to reach $90\%$ of the total counts \\
				& Various functional fits of the mean energy \& counts evolution\\
	GW signal & Various functional fits of the peak frequency evolution\\
					& Timescale to reach 90\% of the maximum peak frequency \\
					& Peak frequency at various times post-bounce \\
					
 \bottomrule
     \end{tabular}
   \label{tab:corr_parameters}
\end{table}

\end{document}